\documentclass[12pt,prc]{revtex4}
\usepackage{amsmath}
\usepackage{epsfig}

\newcommand{\bfsi}{{\mbox{\boldmath$\sigma$}}}

\begin{document}
\title{The role of the in-medium four-quark condensates revised
}
\author{E. G. Drukarev, M. G. Ryskin, V. A. Sadovnikova\\
Petersburg Nuclear Physics Institute NRC ``Kurchatov Institute"}

\begin{abstract}
We calculate the nucleon self-energies in nuclear matter in the QCD sum rules approach, taking into account the contributions of the four-quark condensates. We analyze the dependence of the results on the model employed for the calculation of the condensates and demonstrate that the relativistic character of the models is important. The condensates are calculated with inclusion of the most important terms beyond the gas approximation. This corresponds to inclusion of the two-body nucleon forces and of the most important three-body forces. The results are consistent with the convergence of the operator product expansion. The density dependence of the nucleon self-energies is obtained.
The results are consistent with those obtained by the standard nuclear physics methods, thus inspiring further development of the approach.
\end{abstract}
\maketitle

\section{Introduction}

Calculation of the density dependence of the nucleon self energy is one
of the most important problems of nowadays nuclear physics \cite{A1,
A2}. The standard nuclear physics calculations carried out in various
approaches \cite{A2,A3} require certain phenomenological parameters.
The QCD sum rules approach to description of the nucleons in nuclear matter \cite{1}
enables to present the nucleon self-energies in terms of the in-medium QCD condensates \cite{10}.

The QCD sum rules approach stimulated investigation of the in-medium four-quark condensates.
It was demonstrated \cite{2} that the influence of the
four-quark condensates may appear to be large. However the
calculations require certain model assumption on the quark structure
of nucleus.

In \cite{3} the contribution of the scalar condensate to the QCD sum
rules was obtained by using the Nambu--Jona--Lasinio model \cite{4}. A
larger number of condensates was obtained in \cite{5} in framework of
the Perturbative Chiral Quark Model (PCQM) \cite{6,7}. Further studies
showed \cite{8,9} that a more detailed analysis is needed. Also, the
status of the factorization assumption \cite{10} should be clarified.

In the present paper we consider the four-quark condensates in the
class of models, where the nucleon is viewed as a system of the valence quarks and the pions, which are treated perturbatively. The sea quarks are assumed to be contained in the pions. We focus on the role of the four-quark condensates
in the QCD sum rules
analysis of the nucleons in nuclear matter.

The QCD sum rules approach to investigation of hadrons in vacuum was
suggested in \cite{11} for the mesons and was extended for nucleons in
\cite{12} (see also \cite{13}). The aim of the approach is to express hadron parameters in terms of the QCD condensates. It was extended for calculations of the
nucleon self-energies in
nuclear matter (see \cite{14} and references therein). The method is
based on dispersion relations for the function $\Pi_m(q^2)$ usually referred to as the
``polarization operator"  which
describes the time-space propagation of the system with the four-momentum $q$ and the quantum
numbers of the nucleon in the matter. The function $\Pi_m(q^2)$ is
determined by the local quark operator $j(x)$. The choice of
$j(x)$ is not unique. We shall use the form suggested in \cite{15} (see
Appendix~A).  We shall focus on the contributions of the four-quark
condensates to the function $\Pi_m(q^2)$.

The calculations are carried out in the gas approximation and with inclusion of the
most important nonlinear terms. This
corresponds to inclusion of the two-body forces and of the three-body forces in the mean-field approximation.

The unknowns of the sum rules equations are the nucleon vector self-energy $\Sigma_V$,
the effective mass $m^*$ (or the scalar self-energy $m^*-m$), the nucleon residue $\lambda^2_m$ and the effective continuum threshold $W_m^2$. The sum rules equations connect these parameters to the QCD condensates at finite density.
Putting $W_m^2$ equal to its vacuum value we find explicit approximate equations which express the nucleon self-energies in terms of the QCD condensates.

There are three types of contributions to the four-quark expectation values.
In the first one only the valence quarks are involved. The second one includes only the sea quarks.
There are also the interference contributions, where  both sea and  valence quarks participate.
The contribution of the valence quarks can be written in a way, which enables to separate special {\em factorized} configurations in which two quark operators act inside the nucleon while two other ones act on the QCD vacuum.

It was shown in \cite{16} that the contributions involving only the sea quarks cancel.
Turning to the valence quarks we find  the factorized terms and also
the {\em internal} terms in which all four operators act inside the nucleon.
In the interference terms two quark operators act on the constituent quark, while the other two act on pion or
connect the pion with vacuum by the PCAC relations.

The four-quark condensates have dimension $d=6$. The factorized terms
are proportional to the product $\langle0|\bar qq|0\rangle\cdot \langle
N|\bar q \Gamma_X q|N\rangle_{val} \cdot \rho$ ($\Gamma_X=I,\gamma_0$)
of the vacuum expectation value  $\langle0|\bar qq|0\rangle$,
contribution of the valence quarks to the nucleon expectation value
$\langle N|\bar q \Gamma_X q|N\rangle_{val}$ and the nucleon density
$\rho$. The internal terms depend on the characteristic nucleon size
$R$. They  scale as $\rho/R^3$.  Since $\langle0|\bar qq|0\rangle
\approx(-241$\,MeV)$^3$, while $R
\approx0.6$\,fm\,$=(328\,$MeV$)^{-1}$, these two values are of the same
order.

We demonstrate that in the scalar channel the factorized terms determine
about $80$ percent of the total contribution.
The internal contributions are as important as the interference terms.
In contrary, in the vector channel the nonfactorized terms play the
crucial role. There is a large cancellation between the factorized and
internal terms.

The results are consistent with the assumption of the convergence of the power series in $1/q^2$ for the polarization operator.

We show that in the gas approximation inclusion of the four-quark
condensates subtract more than $100$\,MeV from the vector self-energy
$\Sigma_V$ and  from the nucleon effective mass $m^*$ at the saturation
value of density.  The effect becomes weaker after inclusion of the
main three-body terms, in which the two pairs of the quark operators
act on two different nucleons. The quantitative results depend on the
value of the nucleon matrix element $\kappa_N=\langle N|\bar uu+\bar
dd|N\rangle$. Anyway, at the phenomenological saturation point the
matter appeared to be bound.

In is instructive to compare the results with those, obtained in framework of the nonrelativistic quark model. The results differ strongly from those found in the PCQM. In particular, the matter appeared to be bound only if the value of $\kappa_N$ is large enough.

Our approach enabled to find the density dependence of the nucleon self-energies. The reasonable results of the paper stimulate further calculations with more complete inclusion of the three-body terms.

\section{General equations}

\subsection{Expression for the four-quark condensates}

We calculate the expectations values of the color-antisymmetric
operators
\begin{equation}
T^{f_1f_2}_{XY}\ =\ :\Big(\bar q^{f_1a}\Gamma_X q^{f_1a'}\cdot
\bar q^{f_2b} \Gamma_Y q^{f_2b'} (\delta_{aa'}\delta_{bb'}
-\delta_{ab'} \delta_{ba'}) \Big):\, .
\label{1}
\end{equation}
Here the colon signs denote the normal ordering of the quark operators
$q^f,f_{1,2}$ stand for the quark flavors, while $a,a',b,b'$ are the
color indices. The basic $4\times4$ matrices $\Gamma_{X,Y}$ acting on
the Lorentz indices of the quark operators are
\begin{eqnarray}
&&
\Gamma_X=I, \quad \Gamma_X=\gamma_5\,, \quad \Gamma_X=\gamma_\mu\,,
\quad \Gamma_X=\gamma_\mu\gamma_5\,,
\nonumber\\
&& \Gamma_X\ =\ \frac i2 (\gamma_\mu\gamma_\nu-\gamma_\nu\gamma_\mu)\
=\ \sigma_{\mu\nu}\, ,
\label{2}
\end{eqnarray}
describing the scalar, pseudoscalar, vector, axial and tensor cases
correspondingly. The operators defined by Eq.~(1) can be written also
as
\begin{equation}
T^{f_1f_2}_{XY}\ =\ :\left(\frac23\, \cdot \bar q^{f_1}\Gamma_X I_c q^{f_1}
\cdot \bar q^{f_2} \Gamma_YI_c q^{f_2} - \frac12\, \bar q^{f_1}\Gamma_X
\lambda^\alpha q^{f_1} \cdot \bar q^{f_2}\Gamma_Y \lambda^\alpha q^{f_2}
\right):\ ,
\label{3}
\end{equation}
where $\lambda^\alpha$ are the basic SU(3) matrices, $I_c$ is the unit
$3\times3$ color matrix. We shall omit $I_c$ further.

\subsection{Sum rules}
\subsubsection{Inclusion of the four-quark condensates}
Considering nuclear matter as a system of $A$ nucleons with momenta $p_i$ we introduce the vector
$$
P\ =\ \frac{\sum p_i}{A}\,.
$$
In the rest frame of the matter
$$
P_\mu\ =\ p_0\delta_{\mu 0}
$$
 with $p_0 \approx m$.

The sum rules are based on the dispersion relations
\begin{equation}
\Pi_m^i(q^2)\ =\ \frac1\pi\int\frac{{\rm Im}\Pi^i_m(k^2)}{k^2-q^2}dk^2
\label{4}
\end{equation}
for the components $\Pi^i_m(q^2)$ of the polarization operator
$$
\Pi_m(q^2)=i\int d^4x e^{i(qx)} \langle M|j(x)\bar j(0)|M\rangle=\hat q \Pi^q_m(q^2)+\hat P \Pi^P_m(q^2)+I\Pi^I_m(q^2).
$$
The left-hand side (LHS) of Eq.\,(\ref{4}) is presented as a power
series of $q^2$. This is known as the operator product expansion (OPE).
The functions Im$\,\Pi^i_m(k^2)$ on the right-hand side (RHS) of
Eq.\,(4) are approximated by the ``pole+continuum" model in which the
lowest pole is written exactly, while the higher states are
approximated by continuum.

The Borel transformed QCD sum rules in nuclear matter take the form
\begin{equation}
{\cal L}^q(M^2, W^2_m)=\lambda_m^2\exp{(-m_m^2/M^2)}; \quad
{\cal L}^I(M^2, W_m^2)=m^*\lambda_m^2\exp{(-m_m^2/M^2)}; \quad
\label{5}
\end{equation}
$$
{\cal L}^P(M^2, W^2_m)=-\Sigma_V\lambda_m^2\exp{(-m_m^2/M^2)}\,.
$$
Here ${\cal L}^i$ are the Borel transforms of the LHS of Eq.\,(4) with
the continuum contributions transferred to the LHS, $m_m$ is the new
position of the nucleon pole \cite{14}.  These equations are expected
to determine the nucleon effective mass $m^*$ and the vector
self-energy $\Sigma_V$, the residue at the nucleon pole $\lambda^2_m$
and the effective continuum threshold $W_m^2$.  The explicit form of
the LHS will be presented later.

The LHS of Eqs. (5) which includes the in-medium condensates can be written as (we omit the dependence on $W_m^2$)
\begin{equation}
{\cal L}^q(M^2)=\sum _n\tilde A_n(M^2); \quad
{\cal L}^I(M^2)=\sum_n \tilde B_n(M^2); \quad
{\cal L}^P(M^2)=\sum_n \tilde P_n(M^2)\,.
\label{5a}
\end{equation}
Here the lower indices $n$ denote the dimensions of the condensates.
Inclusion of the higher dimensions corresponds to inclusion of the
higher terms in the $1/q^2$ expansion of the LHS of Eqs.\,(4). Each
term on the RHS of Eqs.\,(\ref{5a}) is the Borel transform of the
corresponding contribution to the polarization operator $\Pi^i(q^2)$
multiplied by the factor $32\pi^4$.

Earlier Eqs.\,(\ref{5}) were solved for the contributions with $n<6$
with inclusion of the radiative corrections of the order $\alpha_s$.
Now we include also the terms with $n=6$, which are mainly the
four-quark condensates.

The contribution of the four-quark condensates to the polarization operator can be written as
\begin{equation}
(\Pi_m)_{4q}\ =\ (\Pi_0)_{4q}+\frac1{q^2} \sum_{X,Y} \Big(\mu^{XY}H_{XY}+
\tau^{XY}R_{XY}\Big),
\label{7}
\end{equation}
where $(\Pi_0)_{4q}$ is the vacuum term, with
\begin{equation}
H_{XY} = \langle M|T^{uu}_{XY}|M\rangle-\langle 0|T^{uu}_{XY}|0\rangle; \quad
R_{XY} = \langle M|T^{ud}_{XY}|M\rangle-\langle
0|T^{ud}_{XY}|0\rangle\,, \label{8}
\end{equation}
while (see Appendix A)
\begin{equation}
\mu^{XY}\ =\ \frac{\theta_X}{16}\mbox{ Tr
}(\gamma^\alpha\Gamma^X \gamma^\beta\Gamma^Y)\,\gamma_5\gamma_\alpha
\hat q \gamma_\beta\gamma_5 \,, \label{9}
\end{equation}
and
\begin{equation}
\tau^{XY}\ =\ \frac{\theta_X}4\mbox{ Tr }(\gamma^\alpha \hat q
\gamma^\beta\Gamma^X)\,\gamma_5\gamma_\alpha\Gamma^Y\gamma_\beta\gamma_5\,.
\label{10}
\end{equation}
Here $\theta_X=+1$, if $\Gamma^X$ has a vector or tensor structure,
while $\theta_X=-1$ in other cases.

Here we omitted the Lorentz indices of the matrices $\Gamma_X$, $\mu^{XY}$ and $\tau^{XY}$.
Note the way, matrices $\Gamma_X$ with the lower indices in Eqs.
(1)--(3), are related to those with upper indices in Eqs. (\ref{9}) and
(\ref{10}). In   the scalar and pseudoscalar cases $\Gamma^X=\Gamma_X$,
 $\Gamma^X=\gamma^\mu$ and $\Gamma^X=\sigma^{\mu\nu}$ in vector and
tensor cases, $\Gamma^X=\gamma_5\gamma^\mu$ in the axial case.

The traces on the RHS
of Eq.\,(\ref{9}), describing contributions of the $4u$ condensates
obtain nonzero values for all five cases listed in Eq.\,(2), while
$\Gamma^X$ and $\Gamma^Y$ should belong to the same channel. In the
case of $2u2d$ condensates (see Eq.\,(\ref{10})) nonzero values are
provided by the vector and axial condensates and also by the
scalar-vector condensate ($\Gamma^X=\gamma^0$, $\Gamma^Y=I$) and by the
axial-tensor condensate ($\Gamma^X=\gamma^5\gamma^{\lambda}$,
 $\Gamma^Y=\sigma^{\mu\nu}$). As noted in \cite{9}, the latter
contribution was overlooked in \cite{5}.  The matrix element $R_{XY}$
can obtain a nonzero value also in the axial-pseudoscalar channel with
$\Gamma_X=\gamma_0\gamma_5$, $\Gamma_Y=\gamma_5$. However, one can see
that the corresponding contribution is proportional to
$q_{\lambda}\varepsilon^{\lambda \alpha \beta 0}\cdot \sigma_{\alpha
\beta}\gamma_5$. It does not contribute to the structures of the
polarization operator proportional to $\hat q$, $\hat P$ or $I$.

Note that in the scalar and pseudoscalar cases
($\Gamma_{X}=\Gamma_Y=I$, $\gamma_5$) the four-quark condensates are
determined by one function of density in each of the channels
\begin{equation}
H_{XX}=a^{Xuu}_m(\rho), \quad R_{XX}=a^{Xud}_m(\rho)\,.
\label{800}
\end{equation}
The same refers to the vector-scalar and axial-tensor condensates (in the latter case $\Gamma_X=\gamma_{\lambda}\gamma_5$, $\Gamma_Y=\sigma_{\mu\nu}$ and $a_m$ is proportional to the
asymmetric tensor $\varepsilon_{0\mu \nu \lambda}$). The vector condensate with $\Gamma_X=\gamma_{\mu}$,  $\Gamma_Y=\gamma_{\nu}$ can be presented in terms of two functions of $\rho$
\begin{equation}
H_{XY}\ =\ a_m(\rho)g_{\mu \nu}+b_m(\rho)\frac{p_{\mu}p_{\nu}}{m^2}\,,
\label{801}
\end{equation}
with similar equation for $R_{XY}$. The same presentation can be written for the axial condensate. Also, there are two functions in the tensor case with
$\Gamma_X=\sigma_{\mu\nu},\Gamma_Y=\sigma_{\alpha\beta}$, i.e.
\begin{eqnarray}
H_{XY}&=& a_m(\rho)(g_{\mu\alpha}g_{\nu\beta}-g_{\mu\beta}
g_{\nu\alpha})\ +
\label{802}
\\
&& +\ b_m(\rho)\frac{g_{\mu \alpha}p_{\nu}p_{\beta}+g_{\nu\beta}
p_\mu p_{\alpha}-g_{\mu \beta}p_{\nu}p_{\alpha}- g_{\nu \alpha}p_{\mu}
p_{\beta}}{m^2}\,.
\nonumber
\end{eqnarray}

\subsubsection{Gas approximation}

In the gas approximation the matrix elements in nuclear matter are
assumed to be
\begin{equation}
\langle M|X|M\rangle\ =\ \langle0|X|0\rangle +\rho_p\langle
p|X|p\rangle + \rho_n\langle n|X|n\rangle
\label{11}
\end{equation}
for any operator $X$. Here $|p\rangle$ and $|n\rangle$ denote the
unpolarized nucleons in QCD vacuum, $\rho_p$ and $\rho_n$ are the
corresponding densities. In the first term on the RHS
of Eq.~(\ref{11}) all four quark operators act on the vacuum state. In
the gas approximation
\begin{equation}
H_{XY}\ =\ h^p_{XY}\rho_p+h^n_{XY}\rho_n\,;
\quad R_{XY}\ =\ r^p_{XY}\rho_p+r^n_{XY}\rho_n\,,
\label{12}
\end{equation}
with the matrix elements for unpolarized nucleons
\begin{equation}
h^N_{XY}\ =\ \langle\,N|T^{uu}_{XY}|N\,\rangle\,; \quad
r^N_{XY}\ =\ \langle\,N|T^{ud}_{XY}|N\,\rangle\,.
\label{13}
\end{equation}
In the gas approximation the functions $a_m(\rho)$ and $b_m(\rho)$
introduced in the previous Subsubsection -- see
Eqs.\,(\ref{800})--(\ref{802}) are proportional to $\rho$. Thus the ratios
$a_m(\rho)/\rho $ and $b_m(\rho)/\rho $ do not depend on $\rho$ and can
be expressed in terms of the nucleon matrix elements $h_{XY}$ and
$r_{XY}$ introduced in Eq.\,(\ref{13}).

In the gas approximation we can write Eq.~(\ref{7}) in the form
\begin{equation}
(\Pi_m)_{4q} - (\Pi_0)_{4q} =\left (A^q\frac{\hat q}{q^2}
 +A^P\frac{(Pq)}{m^2}\,\frac{\hat P}{q^2}
 +A^I\frac{(Pq)}{m}\,\frac{I}{q^2} \right)
 \frac a{(2\pi)^2}\,\rho\,,
 \label{11a}
 \end{equation}
 with
 $$
A^i=\sum A^{i,f_1f_2}_{XY}; \quad i=q,P,I,
$$
where the RHS is summed over the flavors $f_1,f_2$ and the Lorentz
structures $X,Y$. Here we used the standard notation
 $a=-(2\pi)^2\langle0|\bar qq|0\rangle$ $(a>0)$. Note that this is just
 a convenient scale for presentation of the results. It does not
 reflect the chiral properties of the nucleon. The coefficients $A^i$ contain also dependence on
 the asymmetry parameter
 \begin{equation}
 \alpha\ =\ (\rho_n-\rho_p)\Big/(\rho_n+\rho_p).
 \label{100}
 \end{equation}
 Since in the case
 of $4u$ quarks only structures with $X=Y$ contribute, we put
 $A^{uu}_{XX}=A^{uu}_X$.

\subsection{Three types of terms}

Now we turn to the models in which the nucleon is considered as a
 composition of the constituent quark and the pion cloud. The nucleon
 vector of state is thus $|N\rangle=|\phi_0, \pi\rangle$, with $|\phi_0
 \rangle$ the state of three valence quarks. Note that all ingredients
 are in the QCD vacuum. If the pion cloud is treated perturbatively,
the nucleon expectation value of the operator defined by Eq.\,(\ref{3})
can be written as \cite{5}
\begin{equation}
\langle N|T^{f_1f_2}_{XY}|N\rangle=\langle \phi_0|T^{f_1f_2}_{XY}|\phi_0\rangle+
\langle \phi_0|H_I|\phi_0, \pi\rangle\langle \pi|T^{f_1f_2}_{XY}|\pi\rangle \langle \phi_0, \pi|H_I|\phi_0\rangle
+\langle N|T^{f_1f_2}_{XY}|N\rangle_{INTERF}.
\label {15a}
\end{equation}
Here $H_I$ is the interaction between the constituent quarks and the pions.
In the first term on the RHS of Eq.\,(\ref{15a}) all four
 operators act on the constituent quarks. In the second term the four
quark operators act on pions -- Figs. 1 and 2.

In the last term two quark operators act on the constituent quarks.
Two other ones act on the pions. This contact interference is
illustrated by Fig.\,3a. In the vertex interference shown in Fig.\,3b
 the $\pi Q$ interaction $H_I$ adds a pion to the constituent quark. In
 the four-quark operator two operators act on the constituent quarks
 while the two other ones annihilate the pion. In the case of $4u$
 operators we can write
\begin{eqnarray}
&& \hspace*{-0.5cm}\langle N|T^{uu}_{X}|N\rangle_{INTERF}\,=\, \frac43
\,\Big(\langle\phi_0|H_I|\phi_0, \pi\rangle\langle\pi|t_X|\pi\rangle
 \langle \phi_0|t_X |\phi_0\rangle\langle\phi_0,\pi|H_I|\phi_0\rangle+
\label{15b}
\\
&&+\ \langle \phi_0|t_{X}|\phi_0\rangle \langle 0|t_X|\pi \rangle
 \langle \phi_0, \pi|H_I|\phi_0\rangle +\langle \phi_0|H_I|\phi_0,
\pi\rangle \langle\pi|t_X|0\rangle \langle\phi_0|t_X|\phi_0\rangle\Big),
 \nonumber
 \end{eqnarray}
with $t_X=\bar u\Gamma_X I_c u$. Since the vacuum and the pion states
 are colorless, only the first term on the RHS of Eq.\,(3) contributes.
 The first term on the RHS has a nonzero value in the scalar case
 $\Gamma_X=I$.  The two next terms obtain nonzero values in the axial
and pseudoscalar cases with the matrix elements $\langle 0|t_X|\pi
\rangle$ determined by PCAC. Similar equation can be written for $2u2d$
condensate.

\subsection{Contributions to the polarization operator}

Here we consider only those of the four-quark condensates which
contribute to the polarization operator.

\subsubsection{Sea quarks}

In the gas approximation the sea quarks are contained in  the pion clouds of
separate nucleons. If the pions are treated perturbatively, the four
quark operators should act on the same pion.

This requires calculation of the matrix elements
$\langle\pi^\alpha|T^{f_1f_2}_{XY}|\pi^\alpha\rangle$ with
$T^{f_1f_2}_{XY}$ defined by Eqs. (\ref{1}) and (\ref{3}), here $\alpha$
denotes the pion isotopic states. These expectation values can be
expressed in terms of the vacuum expectation values
$\langle0|T^{f_1f_2}_{XY}|0\rangle$ by means of the current algebra
technique\,\cite{18}. If the factorization hypothesis for the latter
expectation values is assumed \cite{11}, one finds that they obtain
nonzero values only for $X=Y$ and \cite{5}
\begin{equation}
\label{16}
\sum_X\Big(\mu_X\langle\pi^\alpha|T^{uu}_{XX}\,|\pi^\alpha\rangle
+\tau_X\langle\pi^\alpha|T^{ud}_{XX}\,|\pi^\alpha\rangle\Big)\ =\ 0
\end{equation}
for any isotope index $\alpha$.

Thus using Eq. (\ref{16}) we find that the contribution of the sea
quarks vanishes.

\subsubsection{Contributions of the valence
quarks. General equations}

The valence quarks are assumed to be represented by the constituent
quarks, i.e. by the massive particles moving in a certain
self-consistent effective field. The constituent quarks are assumed to
be described by the single-particle functions, which are the solutions
of the wave equation. In the general case the constituent quarks are
relativistic particles, described by the wave function
\begin{equation}
\label{17}
\psi_q({\bf r})\ =\ \omega_q f_q(r)
\end{equation}
with $f$ a certain function of $r=|{\bf r}|$ while in
\begin{equation}
\omega_q\ = \left( \begin{array}{c}
\chi_q \\
-i\beta_q(\mbox{\boldmath$\sigma\nabla$})\,\chi_q \end{array} \right)
\label{18}
\end{equation}
$\chi_q$ is the two-component spinor with $\beta_q$ a certain
function of $r$. In the nonrelativistic models $\beta_q=0$.

The QCD operator $q(r)$ acts on the valence quark $|Q\rangle$ placed
into the QCD vacuum:
\begin{equation}
q^a|Q^a(r)\rangle\ =\ \psi_q(r)|Q^a\rangle+q^a(r)|0\rangle \,,
\label{300}
\end{equation}
with $\psi_q(r)$ defined by Eq.~(\ref{17}). The product of the wave
functions describes thus the excess of the quark density inside the
consistent quark over that of vacuum
\begin{equation}
\bar\psi_q(r)\,\psi_q(r)\ =\ \langle Q|\bar q(r)\,q(r)|Q\rangle
-\langle0|\bar qq|0\rangle
\label{19x}
\end{equation}
(recall that $\bar qq=\Sigma\, \bar q^aq^a$).

\subsubsection{Nucleon matrix elements}

Here we present the model-independent relations for the nucleon matrix elements.
The nucleon matrix element for any quark operator $X(r)$ can be written
as \cite{17}
\begin{equation}
\langle\,N|X|N\,\rangle\ =\ \langle\, N|\int d^3r\Big(X(r)-\langle
0|X|0\,\rangle\Big) |N\,\rangle\,,
\label{11b}
\end{equation}
thus expressing the excess of the quark density inside the nucleon over
that in the QCD vacuum. For the scalar two-quark operator this leads to
\[
\langle N|\bar qq|N\rangle\ =\ \langle N| \int d^3r\Big(\bar q(r)q(r)-
\langle0|\bar qq|0\rangle\Big) |N\rangle\,,
\]
while for the other operators $\bar q\Gamma_Xq$ we have just
\[
\langle N|\bar q\Gamma_Xq|N\rangle\ =\ \langle N|\int d^3r\bar q(r)
\Gamma_X q(r)|N\rangle\,,
\]
with most of these expectation values vanishing for the unpolarized
nucleons.

For the four-quark operators $\bar q\Gamma_Xq\bar q\Gamma_Yq$ with
$\Gamma_X=I$, $\Gamma_Y\neq I$, i.e. for the operators
$X=\bar qq\bar q\Gamma_Yq$ one can write Eq.~(\ref{11b}) in another form
\begin{eqnarray}
&& \hspace*{-0.7cm}
\langle N|\bar qq\bar q\Gamma_Y\,q|N\rangle\ =\
2\,\langle0|\bar qq|0\rangle\,\langle N|\bar q\Gamma_Y\,q|N\rangle\ +
\nonumber\\
&& +\ \langle N|\int d^3r\Big(\bar q(r)q(r)-
\langle0|\bar qq|0\rangle\Big)\bar q(r)\Gamma_Y\,q(r)|N\rangle\,,
\label{13a}
\end{eqnarray}
while for the four-quark scalar condensate
\begin{eqnarray}
&& \hspace*{-0.7cm}
\langle N|\bar qq\,\bar q q|N\rangle\ =\
2\,\langle0|\,\bar qq\,|0\rangle\,\langle N|\,\bar qq\,|N\rangle\ +
\nonumber\\
&&+\ \langle N|\int d^3r\Big(\bar q(r)\,q(r)-
\langle0|\,\bar qq\,|0\rangle\Big)^2|N\rangle\,.
\label{14}
\end{eqnarray}
For the other four-quark condensates
$\bar q\tilde\Gamma_X\,q\bar q\tilde\Gamma_Y\,q$ with
$\tilde\Gamma_{X(Y)}=\Gamma_{X(Y)}$ or $\tilde\Gamma_{X(Y)}=\Gamma_{X(Y)}\lambda^\alpha$
(see Eq.~(\ref{3})) we have just
\begin{equation}
\langle N|\,\bar q\tilde\Gamma_X\,q\bar q\tilde\Gamma_Yq\,|N\rangle\ =\
\langle N|\int d^3r\bar q(r)\tilde\Gamma_X\,q(r)\bar q(r)\tilde\Gamma_Y
\,q(r)\,|N\rangle\,.
\label{15}
\end{equation}
We shall see that Eqs. (\ref{13a}), (\ref{14}) will be useful in the
calculation of the contribution of valence quarks.

\subsubsection{Factorized and internal terms}

In our QCD sum rules analysis there are the factorized terms in the
cases of the scalar--scalar $4u$ condensate and of the vector-scalar
$2u2d$  condensate.  Two quark operators act on the QCD vacuum while
two other ones act on the valence quarks of the nucleon. We can write
\begin{eqnarray}
\langle N|\bar uI_c u\bar u I_c u|N\rangle_{val} &=&
\langle N|\bar uI_c u\bar u I_c u|N\rangle_{fact}
+ \langle N|\bar uI_c u\bar u I_c u|N \rangle_{int}\,,
\nonumber\\
 \langle N|\bar u \gamma_0 I_c u\bar d I_c d|N\rangle_{val} &=&
\langle N|\bar u \gamma_0 I_c u\bar d I_cd|N\rangle_{fact}
+ \langle N|\bar u \gamma_0I_c u \bar d I_c d|N\rangle_{int}\
\label{19}
\end{eqnarray}
-- see Fig. 4.
Here the lower index $val$ denotes that we consider the contribution of
the valence quarks. Index $int$ shows that all four quark operators act
on the valence quarks. The scalar factorized term is
\begin{equation}
\label{21}
\langle N|\bar u I_cu\bar u I_cu|N\rangle_{fact}\ =\ 2\langle 0|
\bar uu|0\rangle J\,,
\end{equation}
where
\begin{equation}
\label{21a}
J\ =\int d^3r\,\bar\psi_u(r)\,\psi_u(r)\,,
\end{equation}
is just the contribution of one valence $u$ quark to the expectation value $\langle p|\bar u u|p\rangle$.
Another factorized term is
\begin{equation}
\langle N|\bar u\gamma_0I_c u\bar dd|N\rangle_{fact}\ =\langle0|\bar dd|0\rangle\langle N|\bar u\gamma_0I_c u|N\rangle=
\langle0|\bar dd|0\rangle\cdot n_u\,.
\label{22}
\end{equation}
Here $n_u$ denotes the number of $u$ quarks in the nucleon.
Note that the value of the  condensate $\langle N|\bar u\gamma_0I_c u\bar dd|N\rangle_{fact}$
does not depend on the particular model of the nucleon.

For the other condensates
\begin{equation}
\langle N|\bar u\tilde\Gamma_X u\bar u\tilde\Gamma_Y u|N\rangle_{val}\ =\
\langle N|\bar u\tilde\Gamma_X u\bar u \tilde\Gamma_Y u|N\rangle_{int}.
\label{23}
\end{equation}

The internal terms vanish if a $4u$ operator is
averaged over neutron \cite{5}. For the proton
\begin{equation}
\langle p|T^{uu}_{XX}|p\rangle_{int}\ =\ 2\int
d^3r\bar\Psi_{uu}\Gamma^{(1)}_X \Gamma^{(2)}_X\Psi_{uu}\,,
\label{26}
\end{equation}
with $\Psi_{uu}$ the wave function of the two constituent $u$ quarks in the proton,
the upper indices $(1),(2)$ label the two $u$ quarks. Since the asymmetry of the function $\Psi_{uu}$ is provided by
that of the colors, the spin function is symmetric and the total spin of the two $u$ quarks is $S=1$.
Thus we must put
$(\bfsi_u^{(1)}\cdot\bfsi_u^{(2)})=1$ in the averaging over the
spin states in Eq.~(\ref{26}).  For the $2u2d$ condensates
\begin{equation}
\langle N|T^{ud}_{XY}|N\rangle_{int}\ =\ \sum_q\int
d^3r\bar\Psi_{ud}\Gamma^{(1)}_X \Gamma^{(2)}_Y\Psi_{ud}\,,
\label{26a}
\end{equation}
with $\Psi_{ud}$ the two-particle wave function of a system of $u$ and
$d$ constituent quarks. The sum is taken over all $ud$ states in the
nucleon. Since the spin of the nucleon is $1/2$ while the sum of the
spins of the two $u$ quarks in the proton (and of the two $d$ quarks in
the neutron) is $1$, we must put $\sum_q(\bfsi_u\cdot\bfsi_d)=-4$
in the averaging over the spin states in Eq.~(\ref{26a}).

\subsubsection{Contact and vertex interference}

Start with the $4u$ condensate.
The first term on the RHS of Eq.\,(\ref{15b}) describes the contact
interference, in which two operators act on the constituent quarks
while the two other ones act inside the neutral pions. There is a
nonzero expectation value only for the scalar condensate
$$
\langle\pi|\bar q q|\pi\rangle\ =\ \frac{2m_{\pi}^2}{m_u+m_d}\,,
$$
with $m_{\pi}$,
$m_u$ and $m_d$ the masses of pion and of the light quarks.

Similar situation takes place for the $2u2d$ condensate. Now the
charged pions also contribute. Some of the vector condensates (e.g.
$\langle \pi^0|\bar d \gamma_0 u|\pi\rangle$) have nonzero values.
However, such expectation values do not contain large factor
$m_{\pi}/(m_u+m_d)$. Thus their contribution is much smaller than that
of the scalar term.

In the second term on the RHS of Eq.\,(\ref{15b}) one of the vertices
of interaction between the consistent $U$ quark and neutral pion is
replaced by the four-quark condensate. This will be referred to as the
vertex interference. Nonzero contributions are provided by the axial
and pseudoscalar condensates.

In the case of the $2u2d$ condensates Eq.\,(7) presents the
polarization operator in terms of the four-quark operators $\bar u
\Gamma_X u\bar d \Gamma_Y d$. However in the calculations of vertex
interference involving the charged pions $\pi^{\pm}$ it is more
convenient to present the polarization operators in terms of the
operators $\bar u \Gamma_X d\bar d \Gamma_Y u$. This is because in the
axial and pseudoscalar channels the operators $\bar u \Gamma_X d$ and
$\bar d \Gamma_Y u$ ($\Gamma_{X(Y)}=\gamma_{\mu}\gamma_5, \gamma_5$)
connect pions and vacuum by the PCAC relations. The term
$\sum_{X,Y}\tau^{XY}R_{XY}/q^2$ on the RHS of Eq.\,(7) is replaced by
(see Appendix~A)
\begin{equation}
(\Pi_m)'_{4q}\ =\ \frac1{q^2} \sum_{X,Y}
\xi^{XY}\tilde R_{XY}\,
\label {26c}
\end{equation}
with
\begin{equation}
\tilde R_{XY}\ =\ \langle M|\tilde T^{ud}_{XY}|M\rangle\,,
\label{26d}
\end{equation}
where $\tilde T^{ud}$ is written in terms of operators $\bar u \Gamma_X d\bar d \Gamma_Y u$, and
\begin{equation}
\xi^{XY}\ =\ \frac14\gamma_5\gamma^{\mu}\Gamma^X\gamma^{\nu}\hat
q\gamma_{\mu}\Gamma^Y\gamma_{\nu}\gamma_5\,.
\label{26e}
\end{equation}

\section{Contributions of the four-quark condensates in PCQM}

\subsection{Wave functions}

In the PCQM the single-particle wave functions
defined by Eq.\,(\ref {17}) are
\begin{equation}
f_u(r)\ =f_d(r)\, =\, f(r)\, =\, Ne^{-r^2/2R^2}
\label{27}
\end{equation}
with $N$ determined by the normalization condition
$\int\!\bar\psi_q(r)\gamma_0\psi_q(r)d^3r=1$.
The two particle wave functions take the form of the products
of the wave functions given by Eqs. (\ref{17}) and (\ref{27}) with the
spinors $\chi_{1,2}$ composing the corresponding spin state.  The
functions $\beta_q$ are replaced by constants
\begin{equation}
\beta_u\,=\,\beta_d\,=\,\beta\,=\,0.39\,,
\label{28}
\end{equation}
fitted to reproduce the value of the axial coupling constant.
The parameter $R$ is related to the contribution of the valence quark
core $\langle r_{core}^2 \rangle$ to the proton charge radius $\langle
r^2 \rangle$ \cite{7}.

In the PCQM calculations of the proton charge radius \cite{17a} the value
\begin{equation}
R\ =\ (0.618 \pm 0.006)\mbox{ fm},
\label{28a}
\end{equation}
provides the exact fit for the nowadays value $\langle
r^2\rangle^{1/2}= (0.8768\pm 0.0069)$fm \cite{17b}.

A characteristic size of the internal expectation values is
\begin{equation}
{\cal N}^2\ =\ \int f^4(r)d^3r\ =\ \frac1{(2\pi)^{3/2}}\cdot
\frac1{R^3}\,\frac1{(1+3/2\,\beta^2)^2}\,,
\label{29}
\end{equation}
with the numerical value for $R=0.618$\,fm
\begin{equation}
{\cal N}^2\ =\ 1.37\cdot10^{-3}\,\mbox{GeV}^3\ =\
0.098\,\varepsilon^3_0\,.  \label{30}
\end{equation}
Here
\begin{equation} \varepsilon^3_0\
=\ -\langle0|\bar qq|0\rangle\,, \qquad \varepsilon_0\ =\ 241\,\rm MeV,
\label{31}
\end{equation}
thus the second factor on the RHS of Eq.(17) is  $a/(2\pi)^2=\varepsilon_0^3$.

\subsection{Contributions of the valence quarks}

\subsubsection{Factorized terms}

Employing  the PCQM value $J=0.54$ we find for the factorized term
provided by Eq.\,(\ref{21})
\begin{equation}
A^q_S\ =\ 1.08\left(1-\frac{\alpha}{3}\right),
\label{32}
\end{equation}
with the asymmetry parameter $\alpha$ defined by Eq. (\ref{100}).
For another factorized term expressed by Eq.\,(\ref {22}) we obtain
\begin{equation}
A^I_{VS}\ =\ 2.00\left(1-\frac{\alpha}{3}\right).
\label{42}
\end{equation}
Inclusion of nonlocality of the vector condensate and of some other small corrections \cite{5}
changes the last equation to (see Appendix~B)
\begin{equation}
A^I_{VS}\ =\ 1.14 \left(1-\frac{\alpha}{3}\right).
\label{42v}
\end{equation}

\subsubsection{Internal terms}

The internal contributions of the valence quarks to the parameters
 $A^{q,P,I}$ defined by Eq.\,(\ref{11a}) can be obtained by
straightforward calculations (see Appendix~C). We find the numerical
values for the $4u$ case \cite{5}
\begin{eqnarray}
&& A^q_S=-0.05+0.04\alpha; \quad  A^q_{Ps} = -0.01+0.01\alpha; \quad
A^q_V =0.01(1-\alpha);
\nonumber\\
&& A^q_A = -0.03(1-\alpha);
\quad A^q_T =0.02(1-\alpha)\,,
\label{34}
\end{eqnarray}
and
\begin{equation}
A^P_S =A^P_{Ps} =0; \quad A^P_V =-0.13(1-\alpha); \quad
A^P_A = 0.03(1-\alpha); \quad A^P_T =-0.09(1-\alpha),
\label{35}
\end{equation}
while $A^I=0$ for all these structures.

In the case of the $2u2d$ condensate the nonvanishing contributions are
\begin{equation}
A^q_V\ =\ -0.80\,; \qquad A^q_A\ =\ -0.49\,,
\label{45}
\end{equation}
while
\begin{equation}
A^P_V\ =\ -0.41\,, \quad A^P_A=-0.25,
\label{46}
\end{equation}
and
\begin{equation}
A^I_{VS}\ =\ -0.34\,, \quad  A^I_{AT}\ =\ 0.77\, .
\label{47}
\end{equation}
Thus the total contribution of the internal terms is
\begin{equation}
A^q=-1.35+0.05\alpha, \quad A^P=-0.85+0.19\alpha,
\label{47a}
\end{equation}
$$
A^I\ =\ 0.43\,.
$$

\subsubsection{Total contribution of the valence quarks}

Summing up the contributions provided by Eqs. (\ref{32}), (\ref{42v})
and (\ref{47a}) we find for the total contribution of the valence quarks
\begin{equation}
A^q=-0.27-0.31\alpha, \quad A^P=-0.85+0.19\alpha,
\label{41}
\end{equation}
$$
A^I=1.57-0.38 \alpha.
$$
One can see that there is a strong cancelation between the factorized
and internal terms in the structure $A^q$. Internal terms provide a
small correction to the structure $A^I$.

\subsection {Interference terms}

We now turn to the case in which one of the operators $\bar
q\Gamma^{X,Y}q$ acts on the constituent quark while the other one acts
on the pion. In the PCQM the pions manifest themselves in the
$\pi^{\alpha}Q$ self-energy loops $\Sigma^{\alpha}$ and in the
corresponding exchange terms $I^{\alpha}$ -- see Fig.\,3. They are
related as \cite{5}
\begin{equation}
I^\alpha\ =\ \frac{10}{9}\,\Sigma^{\alpha}\,,
\end{equation}
with the same relation for the contributions of the four-quark condensates.

For the contact interference described by the first term on the RHS of
Eq.\,(\ref{15b})  we find \cite{5} (see Appendix~D)
\begin{equation}
A^q=-0.02+0.01\alpha; \quad A^I=-0.13+0.06\alpha.
\end{equation}

For the vertex interference, described by the second and third terms on
the RHS of Eq.\,(\ref{15b}) we obtain in the case of $4u$ condensate
\begin{equation}
\langle Q|T^{uu}_{XX}|Q, \pi\rangle\ =\ \frac43\,
\langle Q|\bar u\Gamma_{X} u|Q\rangle \langle 0|\bar u\Gamma_X u|\pi
\rangle\,. \label{15bb}
\end{equation}

The matrix elements $\langle 0|\bar q\Gamma_{X}q|\pi\rangle$ obtain
nonzero values for the axial and pseudoscalar operators. Similar
equation can be written for the $2u2d$ condensate. Since the $\pi QQ$
vertex is proportional to the product $(\mbox{\boldmath$\sigma$} \cdot
{\bf k})$, where {\bf k} is the pion momentum, the vertex
interference can take place if the product  $\langle 0|\bar
q\Gamma_{X}q|\pi\rangle \langle Q|\bar q\Gamma_{Y}q|Q\rangle$ also has
this structure.

Using the definitions
\begin{equation}
j_\mu^q=\bar q\gamma_\mu\gamma_5q\,; \quad j_\mu^+=\bar
u\gamma_\mu\gamma_5 d\,; \quad j_\mu^-=\bar d\gamma_\mu\gamma_5u\,,
\label{58}
\end{equation}
one can write the PCAC relations
\begin{equation}
\langle 0|j_{\mu}^{u}|\pi^0(k)\rangle=if_{\pi}k_{\mu},\quad  \langle
0|j_{\mu}^{d}|\pi^0(k)\rangle=-if_{\pi}k_{\mu}; \quad \langle
0|j_{\mu}^{-}|\pi^+(k)\rangle=i\sqrt{2}f_{\pi}k_{\mu}\,,
\label {59}
\end{equation}
and
\begin{equation}
\langle 0|\bar u\gamma_5
u|\pi^0(k)\rangle=-i\frac{f_{\pi}k^2}{2m_u},\quad \langle 0|\bar
d\gamma_5 d|\pi^0(k)\rangle=i\frac{f_{\pi}k^2}{2m_d}; \quad \langle
0|\bar d\gamma_5 u|\pi^+(k)\rangle=i\frac{f_{\pi}k^2}{m_u+m_d}\,,
\label {60}
\end{equation}
which will be used in calculations of the four-quark condensates.

If both $\Gamma_X$ and $\Gamma_Y$ are the axial matrices, the
contribution of the four-quark condensate can be expressed in terms of
the self-energy loops of the constituent quarks (see Appendix~D). We
find
\begin{equation}
A^q=-0.65+0.02\alpha; \quad A^P=-0.04-0.02\alpha.
\label {61}
\end{equation}
The large value of $A^q$ is provided mostly by the charged pions.

If~~~ $\Gamma^X=\Gamma^Y=\gamma^5$
\begin{equation}
A^q\ =\ 0.84-0.07\alpha\,,
\label {62}
\end{equation}
also with the main contribution provided by the charged pions.

There is also a contribution of the axial $\Gamma^X$ and tensor $\Gamma^Y$. It leads to
\begin{equation}
A^I\ =\ 0.36+0.04\alpha\,.
\label {63}
\end{equation}

Thus for the sum of the interference terms -- see Eqs. (57),
(\ref{61})--(\ref{63}))
\begin{equation}
A^q=0.17-0.04\alpha\,, \quad A^P=-0.03+0.02\alpha\,, \label{41k}
\end{equation}
$$
A^I\ =\ 0.23+0.10 \alpha\,.
$$

\subsection{Total contribution}

The sum of the  values given by Eqs.\,(55) and (\ref{41k}) provides for
the coefficients on the RHS of Eq.\,(\ref{11a})
\begin{equation}
A^q=-0.10-0.34\alpha, \quad A^P=-0.88+0.17\alpha,
\label{41ka}
\end{equation}
$$
A^I\ =\ 1.80-0.28 \alpha\,.
$$

In Table~1, we present the contributions of various Lorentz structures to the matrix elements
$H_{XX}$ defined by Eq.\,(\ref{8}) and to the parameters $A^i$
(Eq.\,(\ref{11a})) for the $4u$ condensates. In Table~2, we show similar
data for the $2u2d$ condensates. In the latter case, all the
contributions  except the vertex interference terms involving the
charged pions are calculated in terms of operators $\bar u \Gamma_X
u\bar d \Gamma_Y d$. The interference terms involving the charged pions
are calculated in terms of the operators $\bar u \Gamma_X d\bar d
\Gamma_Y u$. Note that the data in the Tables~1,2 contain the
contributions of the sea quarks for each Lorentz structure. They cancel
in the sum over all Lorentz structures -- see Eq.\,(21).

\subsection {Calculations in the nonrelativistic quark model}

It is instructive to compare the results with those in the old
nonrelativistic quark model (NRQM).  The nucleon is treated as a system
of three nonrelativistic constituent quarks in a self-consistent
central field.  The wave functions can be taken in the form provided by
Eqs. (\ref{17}), (\ref{18}) and (\ref{27}) with $\beta_q=0$.  We must
equal $\langle r_{core}^2 \rangle$ to the proton charge radius. This
leads to $R=0.72$\,fm.  The NRQM values for the symmetric matter are
\begin{equation}
A^q=1.15, \quad A^P=-0.54, \quad A^I=1.45.
\label{41a}
\end{equation}
The strong deviation between the PCQM and NRQM values for $A^q$ is due
to the difference between the values of the contribution of the valence
quarks to the condensate  $ \langle N|\bar u u|N\rangle $ -- see
Eq.\,(\ref{21a}).  Due to relativistic reduction the PCQM value is
about twice smaller than the NRQM one $J=1$.

\section{Solutions of the sum rules equations}

In this Section, we present equations for symmetric nuclear matter
($\alpha=0$).  Following \cite{1,14}  we consider the
dispersion relations Eq.\,(4) at fixed value of the relativistic pair
energy
$$
s\ =\ (P+q)^2\ =\ 4m^2.
$$
This enables to separate the singularities
connected with the nucleon in the matter from those of the matter
itself.  Now we clarify the terms on the LHS of Eqs.\,(\ref{5}).  The
contributions of continuum are expressed in terms of the functions
$E_i(x)$ with $x=W^2/M^2$
\begin{equation}
E_0(x)=1-e^{-x}; \quad
E_1(x)=1-(1+x)e^{-x}; \quad E_2(x)=1-(1+x+x^2/2)e^{-x}. \label{70a}
\end{equation}
The radiative corrections of the order $\alpha_s$ and $\alpha_s \ln
M^2/\Lambda^2$ to the condensates of dimension $n$ contained in the
structures $A_n$, $B_n$ and $P_n$ -- Eq.\,(\ref{5a}) \cite{19} are
expressed by the factors $\tilde r_{A(B,P)n}$ with $\tilde r_{A(B,P)n}=1$ if the corrections are not
included. We assume the numerical
value $\Lambda=230$\,MeV.  In the one-loop approximation this
corresponds to $\alpha_s(1\,$GeV$^2)=0.475$.  The contribution of the
lowest dimension is \cite{12}
\begin{equation}
\tilde A_0\ =\ M^6 E_2(W^2/M^2)\tilde r_{A0}\,,
\label{70b}
\end{equation}
corresponding to the loop of three free quarks with inclusion of the radiative corrections.

Next come the terms, containing the in-medium condensates of the
lowest dimension $d=3$.
\begin{eqnarray}
\tilde A_3
&=&-8\pi^2\frac{(s-m^2)E_0(W^2/M^2)-M^4E_1(W^2/M^2)}{3m}v(\rho)\tilde r_{A3}\,;
\nonumber\\
\tilde B_3 &=& -4\pi^2 M^4E_1(W^2/M^2)\kappa(\rho)\tilde r_{B3}\,;
\nonumber\\
\label{70c}
\tilde P_3 &=&-\frac{32\pi^2}{3}M^4E_1(W^2/M^2)v(\rho)\tilde r_{P3}\,,
\end{eqnarray}
with the vector and scalar condensates
\begin{equation}
v(\rho) =\langle M|\sum_i \bar q^i\gamma_0 q^i|M\rangle= v_N\rho\, ;
\quad v_N\ =\ 3,
\label{70d1}
\end{equation}
and
\begin{equation}
\kappa(\rho)\ =\ \langle M|\sum_i \bar q^i q^i |M\rangle\,,
\label{70d}
\end{equation}
with the sum over the light quarks.

The nonlocality of the vector condensate and the gluon condensate
compose the contributions of dimension $d=4$.  They are numerically
small, thus we do not display them here. However, we take them into
account in our calculations.  The contributions of dimension $d=6$ come
mostly from the four-quark condensates. In vacuum ($\rho=0$), the only
contribution to the polarization operator is (see Eq.\,(\ref{7}))
\begin{equation}
(\Pi_0)_{4q}\ =\ -\frac{2}{3}\frac{\langle 0|\bar u
u|0\rangle^2}{q^2}\hat q \tilde r_{A6}\,,
\label{70y}
\end{equation}
providing
\begin{equation}
\tilde A_6\ =\ \frac{4}{3} a^2 \tilde r_{A6}\,.
\label{70e}
\end{equation}
Recall that $a=-(2\pi)^2\langle 0|\bar q q|0\rangle >0.$
In nuclear matter we must add the contributions of the nucleons.

\subsection{Gas approximation}

The contribution of the density-dependent part of the four-quark condensates
is given by Eq.~(\ref{11}). Their contributions to the LHS of
Eqs.\,(\ref{5}) are
 \begin{equation}
\label{14a}
\tilde A^g_6=-8\pi^2a A^q\rho\,; \quad \tilde B^g_6=-12\pi^2am
A^I\rho\,; \quad \tilde P_6^g=-12\pi^2am A^P\rho\,,
\end{equation}
with
 the parameters $A^i$ given by Eq.\,(\ref{41ka}). The upper index $g$
denotes that the values are calculated in the gas approximation.

In the gas approximation the scalar condensate determined by
Eq.\,(\ref{70d}) can be written as
\begin{equation}
\kappa(\rho)=\langle 0|\sum_i \bar q^i q^i |0\rangle+\kappa_{\rho};
\quad \kappa_{\rho}=\kappa_N\rho; \quad \kappa_N=\langle N|\sum_i \bar
q^i q^i |N\rangle\,.  \label{70f}
\end{equation}
Parameter $\kappa_N$ can
be expressed in terms of the pion--nucleon sigma term  \cite{cd}
\begin{equation}
\kappa_N\ =\ \frac{2\sigma_{\pi N}}{m_u+m_d}\,,
\label{70g}
\end{equation}
while $\sigma_{\pi N}$ is related to observables \cite{cd1,cd11}.
The experimental data on the value of the sigma term are somewhat
controversial \cite{cd3,cd4}, providing rather an interval of
the possible values
\begin{equation}
7\ \leq\ \kappa_N\ \leq\ 12\,.
\label{70h}
\end{equation}

We present the density dependence of the nucleon parameters in Fig.\,5.
The numerical results at the phenomenological value of the saturation
density $\rho_0=0.17\,$fm$^{-3}$ are given in Table~3. One can see that
at $\kappa_N=8$ inclusion of the four-quark condensates subtracts about
$100\,$MeV from the value of the vector self-energy and about
$170\,$MeV from the value of the effective nucleon mass. Also the
residue of the nucleon pole becomes about $2/3$ of its vacuum value.
The influence of the four-quark condensates on the nucleon parameters
is even larger for $\kappa_N=11$.

Employing Eq. (5) we can write
\begin{equation}
F_V(M^2)\equiv \frac{{\cal L}^P(\rho; M^2, W^2_m)}{{\cal L}^q(\rho; M^2, W^2_m)}=-\Sigma_V(\rho); \quad
F_I(M^2) \equiv \frac{{\cal L}^I(\rho; M^2, W^2_m)}{{\cal L}^q(\rho; M^2, W^2_m)}=m^*(\rho).
\label{105}
\end{equation}
The values which approximate the functions $F_{V,I}(M^2)$ can be
treated as the solutions of the sum rules equations.  Replacing the
effective threshold $W^2_m(\rho)$ by its vacuum value $W_0^2$ we find
the approximate solution of the sum rules equations
\begin{equation}
\frac{{\cal L}^P(\rho; M^2, W^2_0)}{{\cal L}^q(\rho; M^2,
W^2_0)}=-\Sigma_V(\rho)\,; \quad \frac{{\cal L}^I(\rho; M^2,
W^2_0)}{{\cal L}^q(\rho; M^2, W^2_0)}=m^*(\rho)\,, \label{105a}
\end{equation}
which solve the problem of expressing the parameters of the nucleon in
nuclear matter in terms of the density-dependent QCD  condensates.
Putting $W^2_m=W^2_0$ we find $\Sigma_V=253\,$MeV and $m^*=599$\,MeV
for $\rho=\rho_0$ and $\kappa_N=8$, while the solutions of
Eq.\,(\ref{105}) are $\Sigma_V=217\,$MeV and $m^*=562\,$MeV -- see
Table~3.  Hence the accuracy of the approximation $W_m^2=W_0^2$ for the
self-energies $\Sigma_V$ and $m^*\!-\!m$ is about $15$ percent.  It
becomes less accurate for larger values of $\kappa_N$ providing the
errors of about $20$ percent at $\kappa_N=11$.

Note that inclusion of the four-quark condensates lead to
$m^*+\Sigma_V<m$, i.e. the nuclear matter is bound.

The large impact of the four-quark condensates on nucleon parameters
is to some extent caused by employing the gas approximation.

\subsection{Beyond the gas approximation}

Now we take into account the contributions to the four-quark condensates, in which two pairs of quark operators act on two different nucleons of the matter. We include only the contributions containing the large parameter $\kappa_N$.
These are the scalar and scalar-vector contributions.

The contribution proportional to $\kappa_N^2$ can be obtained by
replacing the factor  $(\langle 0|\bar u u|0\rangle)^2$ on the RHS of
Eq.\,(\ref{70y}) by $(\kappa_\rho/2)^2$ -- Eq.\,(\ref{70f}). It comes
from the $4u$ condensate and contributes to the $\hat q$ structure. In
order to obtain the contribution proportional to the product
$\kappa_Nv_N$ it is sufficient to replace the vacuum expectation value
$\langle 0|\bar d d|0\rangle$ in Eq.\,(\ref{22}) by $\kappa_{\rho}/2$.
This contributes to the scalar structure $I$ comes from $2u2d$
condensates.

The corresponding contributions to the matrix elements defined by
Eq.\,(\ref{8}) are
\begin{equation}
H_{SS} =\frac{1}{4}\kappa_N^2\rho^2\,; \qquad
R_{VS} =\frac{1}{4}v_N\kappa_N\rho^2\,.
\label{8b}
\end{equation}

The contributions to the LHS of Eq. (\ref{5}) are now
 \begin{equation}
\label{14n}
 \tilde A_6 =\tilde A^g_6+\frac{16\pi^4}{3}\kappa_N^2\rho^2\,;
  \quad \tilde B_6=\tilde B^g_6+8\pi^4v_N\kappa_N\rho^2\, ;
 \end{equation}
 $$
\tilde P_6\ =\ \tilde P_6^g\,.
$$

Thus the large contribution of the four-quark condensate in the scalar
channel $A^I$ (see Eq.\,(\ref{41ka})), obtained in the gas
approximation becomes about 30 percent smaller due to inclusion of
the 3-body interactions.

The density dependence of the nucleon parameters is shown in Figs. 5
and 6.  The numerical results for $\rho=\rho_0$ are shown in Table~3.
One can see that the influence of the four-quark condensates on the
values of the nucleon effective mass and of the residue becomes weaker,
than it was in the gas approximation.

\subsection{Results in the nonrelativistic quark model}

Now we present the results with contributions of the four-quark
condensates calculated in framework of the NRQM . We find that
for $\kappa_N=8$ they differ strongly from those obtained in the
PCQM -- see Table~4.  The vector self-energy $\Sigma_V$ is about
300\,MeV, and the effective mass $m^*$ differs from the vacuum value
only by about 100--130\,MeV. The nuclear matter appears to be unbound
since $m^*+\Sigma_V>m$. This is due to large factorized term in the $q$
structure which is not compensated by the internal terms in the case of
NRQM. For $\kappa_N=11$ the scalar term increases the value of $m-m^*$
and the matter becomes bound. However the values of the parameters
differ much from those predicted by the PCQM.

\section{Summary}

We calculated the four-quark condensates in nuclear matter in the gas
approximation and included the most important terms beyond it. This
corresponds to inclusion of the two-body forces and of the most
important three-body forces. We applied the results for solving the QCD
sum rules equations (Eq.\,(\ref{5})).

The calculation of the four-quark condensates requires certain model
assumptions on the structure of the nucleons which compose the matter.
In the broad class of models the nucleons are treated as the systems of
valence quarks, described by the constituent quarks and the sea quarks,
contained in  pions. We showed that in the models which treat the pions
perturbatively, there is a remarkable cancelation of the contributions
to the sum rules which contain only the sea quarks. Among the various
contributions which involve the valence quarks one can separate the
factorized terms, where two operators act on the nucleons of the
matter, while two other ones act on the QCD vacuum and internal terms
in which all four operators act on the valence quarks.  In the
interference terms two of the operators act on the valence quarks. The
other two ones act on the sea quarks (i.e. on the pions) or connect the
pions with vacuum by the PCAC relations.

Calculations, carried out in framework of PCQM showed that the
contributions of the four-quark condensates to the scalar channel $A^I$
can be viewed as dominated by the factorized terms. The contribution
to the vector structure $A^P$ is almost totaly due to the internal
terms. The vector structure $A^q$ is a result of strong compensation of
the factorized term given by Eq.\,(\ref{32}) by the internal and
interference terms.

Solving the sum rules we find the nucleon self-energies presented by
Eq.\,(\ref{105}). Approximate solution presented by Eq.\,(\ref{105a})
solves the problem of expressing the nucleon self-energies in terms of
the QCD condensates.

The values of the nucleon parameters at the phenomenological value of
the saturation density are presented in Table~3.  One can see that the
large reduction of the nucleon effective mass $m^*$ and of the nucleon
residue $\lambda_m^2$ is to some extent an artifact of the gas
approximation. The effect becomes less pronounced when the most
important terms beyond the gas approximation are taken into account.

Anyway, the nuclear matter at the phenomenological saturation value
appeared to be bound. The values of the four-quark condensates are
consistent with the hypothesis of the convergence of the power series
in $1/q^2$ of the polarization operator.  Uncertainties in the
experimental data  on the nucleon matrix element $\kappa_N=\langle
N|\bar uu+\bar dd|N\rangle$ (Eq.\,(\ref{70h})) lead to uncertainties in
the quantitative results.

The density dependence of nucleon parameters found by employing the
PCQM are shown in Figs.~5,6. Also, in Fig.~7, we compare the results for
two values of $\kappa_N$.

It is instructive to have a look at the results obtained in the
nonrelativistic quark model. The contribution of the four-quark
condensates to the coefficient $A^q$ differers strongly from that in
PCQM -- Eq.\,(\ref{41a}) since there is no relativistic reduction of
the contribution of the valence quarks to the expectation value
$\langle N|\bar uu |N\rangle$ this time. The results for the nucleon
parameters presented in Table~4 contradict to the nuclear phenomenology
(the matter appears to be not bound) for $\kappa_N=8$. It is bound for
$\kappa_N=11$ but the quantitative results differ from those, obtained
in the PCQM.

The results of further investigation with more complete inclusion of
the many-body forces will be published elsewhere.  The authors
acknowledge the partial support by the RFBR grants 11-02-00120 and
12-02-00158 and by the grant RSGSS -65751.2010.2.

\def\thesection{Appendix \Alph{section}}
\def\theequation{\Alph{section}.\arabic{equation}}
\setcounter{section}{0}

\section{}
\setcounter{equation}{0}

Here  we calculate the contribution of the four-quark condensates to
the polarization operator
\begin{equation}
\Pi_m(q^2)\ =\ i\int d^4x e^{i(qx)} \langle M|Tj(x)\bar j(0)|M\rangle\,.
\label {X1}
\end{equation}

We employ the current suggested in \cite{15}. It is
\begin{equation}
j_\gamma(x)\ =\
\varepsilon^{abc}(u_a^TC\gamma_\mu u_b)(\gamma_5\gamma^\mu d_c)_\gamma\,,
\label {X2}
\end{equation}
where $u$ and $d$ are the quark operators, $a,b,c$ are the color
indices, $C$ is the charge-conjugation matrix $(C^T=C^{-1}=-C)$ and
the upper index $T$ denotes the transposition.  Since we want to
calculate $\Pi_m(q^2)$ as a power series of $q^{-2}$, we need expansion
of the function
\begin{eqnarray}
{\cal F}_{\gamma\zeta}(x)\ \equiv\ T j_{\gamma}(x)\bar j_\zeta(0) &=&
-i\varepsilon^{abc}\varepsilon^{def}u^a_\alpha(x)(C\gamma_\mu)_{\alpha\beta}
u^b_{\beta}(x)(\gamma_5\gamma^{\mu})_{\gamma \delta} d^c_{\delta}(x)\
\times
\nonumber
\\
\label {X3}
&& \times\ \bar d^f_\tau(0)(\gamma^\nu\gamma_5)_{\tau\zeta}\bar
u^e_\eta(0)(\gamma^\nu C)_{\eta\beta}\bar u(0)^d_\beta\,.
\end{eqnarray}
in powers of $x^2$. For a product of the two quark operators we can
write
\begin{eqnarray}
q^a_{f_1\alpha}(x)\bar q^b_{f_2\beta}(0) &=&
G_{\alpha\beta}^{ab}(x)\delta_{f_1f_2}-\frac1{12}\sum_X\bar q^a_{f_1}(0)
\Gamma_X q^b_{f_2}(0)\Gamma^X_{\alpha\beta}\delta_{ab}\ -
\nonumber
\\
\label{X4}
&&-\ \frac18\sum_{X,A}\bar q^a_{f_1}(0)\Gamma_X\lambda^A q^b_{f_2}(0)
\Gamma_{\alpha \beta}^X\lambda^A_{ab}+0(x^2)\,,
\end{eqnarray}
with the fundamental matrices $\Gamma_X$ defined by Eq.\,(2), $f_{i}$
are the quark flavors, while $G^{ab}(x)=G(x)\delta^{ab}$ where
\begin{equation}
G(x)\ =\ \frac{i}{2\pi^2}\frac{\hat x}{x^4}
\label{X5}
\end{equation}
is the propagator of a free massless fermion.

In the case of $4u$ condensates, we describe the product $
d^c_{\delta}(x)\bar d^f_\tau(0)$ in Eq.\,({\ref{X3}) by the first term
on the RHS of (\ref{X4}), while the other products of the field
operators are given by the two next terms on the RHS of that
equation. Since
\begin{equation}
\int d^4 x G(x)e^{i(qx)}\ =\ \frac{i \hat q}{q^2}\,, \label{X6}
\end{equation}
we come to Eq.\,(9) after doing some matrix algebra. We employed that $C\Gamma_X^TC=\pm\Gamma^X$ with the
plus sign in the vector and tensor cases and the minus sign in other
cases.

In the case of $2u2d$ condensates the product of two $u$ quark
operators (e.g.  $u^a_{\alpha}(x)\bar u^e_{\eta}(0)$) is described by
the propagator $G^{ab}(x)$ given by Eq.\,(\ref{X5}). Employing
Eq.\,(\ref{X4}) we describe the product of the operators $u\bar u d\bar
d$ in terms of the operators $\bar u \Gamma_X u\cdot \bar d\Gamma_Y d$,
coming to Eq.\,(10).  An alternative presentation in terms of the
operators $\bar u Z_X d\cdot \bar d Z_Yu$ can be obtained by employing
the commutation relations for the quark operators and presenting the
products $\bar u d$ and $\bar d u$ employing Eq.\,(\ref{X4}). This
leads to Eqs. (\ref{26c})--(\ref{26e}). The two presentations are tied
by the Fierz transform.

\section{}
\setcounter{equation}{0}

The general expression for the contribution of the factorized
vector-scalar term (Eq.(27)) to the LHS of Eq.\,(4) is \cite{16}
\begin{equation}
\Pi_{fact}(q)\ =\ -\frac{2\langle 0|\bar dd|0\rangle}{3}\int\frac{d^4x}{\pi^2x^4}(x \cdot
\theta(x))e^{i(qx)}\rho_p\,,
\end{equation}
with~~ $\theta(x)=\theta^u(x)+\theta^d(x)$, while
\begin{equation}
\theta_\mu^q(x)\,=\,\langle p|\bar q(0)\gamma_\mu q(x)|p\rangle\,
=\,\frac{p_\mu}{m}\varphi_a ((px),x^2)+ix_\mu
m\varphi_b((px),x^2)\,, \label{B1}
\end{equation}
where
\begin{equation}
q(x)\ =\ q(0)+ x_{\alpha}D^{\alpha}q(0)+...
\end{equation}
In the leading order we can put $x^2=0$ in Eq.\,(\ref{B1}) and define
\cite{16,14}
\begin{equation}
 \varphi_{a(b)}((px))\ \equiv\
\varphi_{a(b)}((px,0))\ =\int_0^1d\alpha
f_{a(b)}(\alpha)e^{-i(px)\alpha}\,.
 \end{equation}
 with
$f_a(\alpha)=f(\alpha)$ the proton deep inelastic structure functions
\cite{20}. The moments of the functions $f_b(\alpha)$ can be expressed
in the terms of the moments of the functions $f(\alpha)$ \cite{14}.
Including the two lowest moments $\langle\alpha^{n-1} f\rangle$ we find
 \begin{equation}
\Pi(q)_{fact}\ =\ -\frac{2\langle 0|\bar d d|0\rangle}{3q^2}
 \Big(\frac{2(pq)}{m}\langle f\rangle- \frac{4(pq)}{m}\langle\alpha
f\rangle-m\langle\alpha f\rangle \Big)\rho_p\,.
\end{equation}
Using
the normalization condition $\langle f\rangle=3$ and the numerical
value $\langle\alpha f\rangle=0.45$ \cite{21} we come to
$A^I_{VS}\ =\ 1.30 \left(1-\alpha/3\right).$ Inclusion of the lowest
correction to the vacuum expectation value $\langle 0|\bar dd|0\rangle$
\cite {16} leads to Eq.\,(\ref{42v}).

\section{}
\setcounter{equation}{0}

Here we give several examples of calculation of the internal terms
-- Eqs.\,(\ref{26}) and (\ref{26a}). For the vector and axial
terms the matrix elements for the $4u$ condensate defined by
Eq.\,(\ref{13}) is
\begin{equation}
\label{A1}
h^p_{\mu\nu}\ =\ a^{V(A)}g_{\mu\nu}+b^{V(A)}
\frac{p_{\mu}p_{\nu}}{m^2}\,,
\end{equation}
with $\mu (\nu)$ the
indices of the vector and axial matrices $\gamma_{\mu}$ and
$\gamma_{\mu}\gamma_5$.  There is similar equation for the $2u2d$
condensate described by the matrix element $r^N_{\mu\nu}$.

In the vector case one can calculate for the time components
\begin{equation}
\label{A2}
h^p_{00}\ =\ a^V+b^V\,=\int d^3x f^2(x)
\Big(1-\beta^2\frac{x^2}{R^2}\Big)^2,
\end{equation}
and for the space components
\begin{equation} \label{A3}
\frac13h^p_{ij}\delta_{ij}\ =\ -a^V\ =\ \frac{4\beta^2}{3}\int d^3x
f^2(x)\frac{x^2}{R^2}w\,,
\end{equation}
with
\begin{equation}
\label{A4}
w\ =\ 2\langle \chi|\bfsi^{(1)}_u \cdot \bfsi^{(2)}_u|\chi
\rangle\,.
\end{equation}
Here $\chi$ is the spin wave function of the
nucleon, while $1$ and $2$ label the two $u$ quarks. Since the total
spin of two $u$ quarks is $S=1$, we find $w=2$.

Similar equations can be written for the $2u2d$ condensate. However
this time
\begin{equation}
\label{A5}
w\ =\ \sum\langle\chi|\bfsi_u\cdot\bfsi_d|\chi\rangle\,,
\end{equation}
with the sum over all quarks of the nucleon. For the proton
 $w=\langle\chi|(\bfsi_u^{(1)}+ \bfsi_u^{(2)})\cdot \bfsi_d|\chi \rangle$.
Since the spin of the proton is $S=1/2$ we
can write $1/4(\bfsi_u^{(1)}+ \bfsi_u^{(2)}+\bfsi_d)^2=3/4$.
Employing $(\bfsi^{(1)}_u \cdot \bfsi^{(2)}_u)=1$ we find $w=-4$.
In the same way we find $w=-4$ for
the neutron as well.

This provides for the $4u$ condensate
\begin{equation}
\label{A6}
a^V =-\frac{4}{3}{\cal N}^2\beta^2= -0.20\varepsilon_0^3; \quad
b^V =2{\cal N}^2\Big(1+\frac{13}6\beta^2
+\frac{15}{16}\beta^4\Big)=0.26\,\varepsilon_0^3\,,
\end{equation}
while for $2u2d$ condensate
\begin{equation}
\label{A7}
a^V = \frac83{\cal N}^2\beta^2=0.04\varepsilon_0^3; \quad
b^V =2{\cal N}^2\Big(1+\frac{1}6\,\beta^2
+\frac{15}{16}\beta^4\Big)=0.21\,\varepsilon_0^3\,.
\end{equation}

One can see that in axial case the time components turn to zero \cite{5}.
Thus
\begin{equation}
\label{A8}
a^{A}+b^{A}\ =\ 0,
\end{equation}

Employing the values of $\mu^{XY}$ and $\tau^{XY}$ (see Eqs. (9) and (10)) we
find the contributions presented by Eqs. (\ref{34})--(\ref{46}).

In the axial-tensor case
$\Gamma^X=\gamma_5\gamma^\lambda(\Gamma_X=\gamma_\lambda\gamma_5),~~
\Gamma^Y=\sigma^{\mu \nu}$, we obtain for the ingredients of
Eq.\,(\ref{7})
\begin{equation}
\mbox{Tr}(\gamma^\alpha\hat
q\gamma^\beta\Gamma^X)=\mbox{Tr} (\gamma^\alpha\hat q\gamma^\beta
\gamma^5\gamma^\lambda)= \mbox{Tr} (\hat
q\gamma^\alpha\gamma^\beta\gamma^\lambda\gamma^5) =
4i\varepsilon^{\kappa\alpha\beta\lambda}q_\kappa\, ,
\label{48}
\end{equation}
while
\begin{equation}
\gamma_5\gamma_\alpha\Gamma^Y\gamma_\beta\gamma_5\ =\
ig^{\mu\mu'}g^{\nu\nu'} (g_{\mu'\alpha}g_{\nu'\beta}
-g_{\mu'\beta}g_{\nu'\alpha})\,,
\label{49}
\end{equation}
and thus
\begin{equation}
\tau^{AT\mu\nu\lambda}\ =\ 2q_\rho\,\varepsilon^{\rho\mu\nu\lambda}\,.
\label{50}
\end{equation}

Employing Eq.\,(\ref{18}) for $\omega_q$ we find that the axial and
tensor matrix elements are \cite{5}
\begin{eqnarray}
&& \bar\omega_q\gamma_0\gamma_5\omega_q=0; \quad \bar \omega_q
\gamma_i\gamma_5\omega_q=\sigma_i+\beta^2 \frac{(\bfsi {\bf x})
\sigma_i(\bfsi{\bf x})}{R^2}\,, \quad  (i=1,2,3), \label{CX}
\\
&& \bar \omega_q \sigma_{0j}\omega_q=-2\beta\frac{x_j}{R}; \quad
\bar \omega_q \sigma_{ij}\omega_q=\varepsilon_{ijk}\Big(\sigma_k-\beta^2
\frac{(\bfsi{\bf x})\sigma_k(\bfsi{\bf x})}{R^2}\Big), \quad  (i=1,2,3),
\end{eqnarray}
where $\varepsilon_{ijk}$ is the standard
three-dimensional asymmetric tensor.

Thus the indices $\mu,\nu$ and $\lambda$ on the RHS of Eq.(\ref{50})
should correspond to the space components, while $\rho=0$. Hence we can write
\begin{equation}
\tau^{AT\mu\nu\lambda}\ =\ 2q_0\varepsilon^{0ijk}\ =\ -2q_0\varepsilon_{0ijk}\, ,
\label{51}
\end{equation}
with $i,j,k$ -- the space components, and $\varepsilon_{0123}=1$. Hence
we can put $\varepsilon_{0ijk} = \varepsilon_{ijk}$.

Thus
\begin{equation}
\tau^{AT\mu\nu\lambda}r^{AT}_{\mu\nu\lambda}\ =\ 2q_0\sum\langle
\chi|\bfsi_u \cdot\bfsi_d|\chi \rangle \int d^3x
f^2(x)\left(1-\frac{15}{16}\beta^4\right).  \label{52}
\end{equation}
Hence
\begin{equation}
\tau^{AT\mu\nu\lambda}r^{AT}_{\mu\nu\lambda}\ =\ 7.83\,q_0{\cal N}^2\,
=\,0.77\,q_0\varepsilon_0^3\,.
\label{A52}
\end{equation}
This leads to Eq.(53)
\begin{equation}
A^I_{AT}\ =\ 0.77.
\label{A53}
\end{equation}

\section{}
\setcounter{equation}{0}

Here we calculate some of the interference terms. The contact
interference and the vertex interference in the axial case are
described in \cite{5} in details. Consider here the pseudoscalar vertex
interference.

Start with the case of the $4u$ condensate. Only the neutral pions
contribute to the interference terms. For the matrix element
$h=h^N_{Ps}$ defined by Eq.\,(\ref{13}) we obtain \cite{5}
\begin{equation}
h\ =\ \frac83\,\frac{19}9\,J^0n_u\,.
\end{equation}
Here $J^0$ is the result of replacement of one of the $UU\pi^0$ vertices
in the self-energy loop of the $U$ quark
\begin{equation}
\Sigma^0\ =\ -\frac1{2f_\pi^2}\int d^3xd^3zg^*(x)S(x)D(x-z)S(z)g(z)\,,
\end{equation}
by the factor $-if_\pi^2 m_\pi^2/2m_u$ -- see Eq.~(\ref{60}),
i.e.
\begin{equation}
J^0=\frac{m_{\pi}^2}{2m_u}I; \quad I=\frac12\int d^3xd^3zg^*(x)
S(x)D(x-z)g(z)\,.
\end{equation}
In these equations
$$
g(x)\ =\ \bar\psi(x)i\gamma_5\psi(x)\,,
$$
$D$ is the pion propagator, while \cite{7}
$$
S(x)\ =\ \frac{1-3\beta^2}{2\beta R}+\frac{\beta}{2R^3}x^2\
$$
is the scalar field.

Using the explicit expressions for the functions $\psi(x)$ we obtain
\begin{equation}
I\ =\ \frac{\sqrt{2}\beta}{4\pi^2(1+3/2\beta^2)^2R^2}\int
dy\frac{y^4\Big(1-\frac{\beta^2}2(1+y^2)\Big)}{y^2+\frac{m_\pi^2R^2}2}
\,e^{-y^2}.
\end{equation}
Direct computation provides $J^0=5.20\cdot 10^{-2}\varepsilon^3.$

Thus we obtain $h=0.27\,n_u$. Since $\mu^{Ps}=\hat q/2$ (Eq.\,(9)), we
find the contribution
\begin{equation}
A^q\ =\ 0.20-0.07\alpha\,.
\end{equation}

In the case of $2u2d$ condensates we write the contribution in terms of
the operators $\bar u\gamma_5 d\bar d\gamma_5u$ -- see Sec.\,2.4.4.
Defining the matrix element in Eq.\,(\ref{26d}) as
$\tilde R_{Ps}=\langle M|\tilde T^{ud}_{Ps}|M\rangle=r\rho$. We obtain
\begin{equation}
r=\frac{4}{3}\frac{19}{9}J^C(n_u+n_d); \quad
J^C=\frac{2m_{\pi}^2}{m_u+m_d}I\,,
\end{equation}
with the numerical value $J^C=7.56\cdot 10^{-2}\varepsilon_0^3$\,,
leading to
 \begin{equation}
A^q\ =\ 0.64\,,
\end{equation}
with the total contribution of the pseudoscalar terms (Eq.\,(63))
\begin{equation}
A^q\ =\ 0.84-0.07\alpha\,.
\end{equation}

Turn now to the axial-tensor interference. Only the $2u2d$ operators
contribute.  Start with the case of neutral pions $\pi^0$.  The
$\pi_0DD$ vertex of the $D$ quark self-energy loop can be replaced by
the axial-tensor condensate
\begin{equation}
t_{\lambda\mu\nu}\ =\ \frac23\langle 0|j_{\lambda}^{u}|\pi^0\rangle
\langle D|\bar d\sigma_{\mu\nu}d|D\rangle\,.
\end{equation}
Since only the space components contribute to polarization operator
(Eq.\,(\ref{CX})), we obtain
\begin{equation}
t_{\ell ij}\ =\ \frac{2}{3}if_{\pi}\varepsilon_{ijr}
\langle D|\bar d k_{\ell}\sigma_{r}d|D\rangle\,,
\end{equation}
with the three-dimensional indices $i,j,\ell$  corresponding to the
four-dimensional ones $\lambda\mu\nu$. Using Eq.\,(\ref{CX}) and
employing $\varepsilon_{ijr}\varepsilon_{ij\ell}=2\delta_{r\ell}$ we
obtain
\begin{equation}
\tau^{AT}_{rij} t_{\ell ij}\ =\ -\sum_{ij}\varepsilon_{rij}t_{\ell ij}\ =\
-\frac43if_\pi\langle D|\bar d({\bf k}\cdot\bfsi)d|D\rangle\,.
\end{equation}
This leads to
\begin{equation}
h=\frac{4}{3}\frac{19}{9}J^{AT}n_d; \quad
J^{AT}=\frac{1}{\beta R}I=2.89\cdot 10^{-2}\varepsilon_0^3\,,
\end{equation}
leading to
\begin{equation}
A^I\ =\ 0.12+0.04\alpha\,.
\end{equation}
Proceeding in the same way in the case of charged pions and using
Eqs. (\ref{26c})--(\ref{26e}) we find
$$
A^I\ =\ \frac{4}{3}\frac{19}{9}J^{AT}(n_u+n_d)\,,
$$
providing
\begin{equation}
A^I\ =\ 0.24\,.
\end{equation}

\clearpage

\newpage

\begin{table}
\caption{Contributions of various Lorentz structures to the matrix
elements $H_{XX}$ defined by Eq.\,(\ref{8}) and to the parameters $A^i$
(Eq.\,(\ref{11a})) for the $4u$ condensates.}
\begin{center}
\begin{tabular}{|c|c|c|c|c|c|} \hline
 X&  $a_m/(\rho \varepsilon_0)$& $b_m/(\rho \varepsilon_0)$& $A^q$ &
$A^P$ & $A^I$\\
\hline

S&$-6.34+0.60\alpha $& -& $3.17-0.30\alpha$& 0& 0\\

Ps& $1.74-0.12\alpha $& -& $-0.87-0.06\alpha$&0& 0\\

V& $-0.55+0.01\alpha$&$0.13-0.13\alpha$ & $0.55-0.01\alpha$ &$-0.13+0.13\alpha$&0\\

A&$ 0.33+0.06\alpha$&$0.07-0.05\alpha$& $ 0.47+0.05\alpha$&$0.07-0.05\alpha $&0\\

T&$0.02-0.02\alpha $& $-0.04+0.04\alpha $ & $0.02-0.02\alpha $& $-0.09+0.09\alpha $&0\\

\hline
\end{tabular} \end{center}
\end{table}

\newpage

 \begin{table}
\caption{Contributions of various Lorentz structures to the matrix elements
$H_{XY}$ and $R_{XY}$ defined by Eq.\,(\ref{8}) and to the parameters
$A^i$ (Eq.\,(\ref{11a})) for the $2u2d$ condensates. All the
contributions except the vertex interference terms involving the
charged pions are calculated in terms of operators $\bar u \Gamma_X
u\bar d \Gamma_Y d$.  They are given in the lines above the horizontal
line. The vertex interference terms involving the charged pions are
calculated in terms of the operators $\bar u \Gamma_X d\bar d \Gamma_Y
u$. Their contributions are presented in the lines below the horizontal
line.}

\begin{center}
\begin{tabular}{|c|c|c|c|c|c|} \hline
XY &  $a_m/(\rho \varepsilon_0)$& $b_m/(\rho \varepsilon_0)$& $A^q$ &
$A^P$ & $A^I$\\
\hline
VV&$0.58$&0.21& $-6.22$& -0.41& 0\\

AA& $-0.22$&-0.16 & $2.60$&-0.32& 0\\

VS& $-0.33+0.16\alpha$&-& 0&0&$0.67-0.32\alpha$\\

AT&$ -0.89-0.04\alpha$&-&0&0&$0.89+0.04\alpha$\\
\hline
PsPs&0.64& -& 0.64&0&0\\
AA&0.23&0 & -0.46&0&0\\
AT&0.24& -&0&0&0.24\\
\hline
\end{tabular} \end{center}
\end{table}

\newpage

 \begin{table}
\caption{Nucleon parameters in symmetric nuclear matter at the
phenomenological saturation value of the nucleon density. The first line
shows the vacuum values. The results marked as $(^*)$ in the second and
fifth lines are for the vacuum values of the four-quark condensates.
The values $(GA)$ in the third and sixth lines correspond to the gas
approximation. The values in the fourth and seventh lines correspond to
inclusion of the three-body interactions, following Eq.\,(\ref{14n}).
The results are presented for two values of the parameter $\kappa_N$ --
Eq.\,(\ref{70h})}.
\begin{center}
 \begin{tabular}{|c|c|c|c|c|} \hline
 &  $\Sigma_v$, MeV & $m^*$, MeV & $\lambda^2_m\rm\ ,GeV^6$ &
$W^2_m\rm\ ,GeV^2$ \\
\hline
$\rho=0$ & 0 & 940 & 1.94 & 2.16\\
\hline
$\rho=\rho_0$, $\kappa_N=8$ (*) & 329& 737 & 2.39 & 2.33\\

$\rho=\rho_0$, $\kappa_N=8$ (GA)& 217& 562& 1.21&1.62\\

 $\rho=\rho_0$, $\kappa_N=8$ & 206 & 602& 1.41& 1.69\\
\hline
$\rho=\rho_0$, $\kappa_N=11$ (*) & 335& 606 & 1.91 & 2.12\\

$\rho=\rho_0$, $\kappa_N=11$ (GA)& 203& 388& 0.98&1.52 \\

 $\rho=\rho_0$, $\kappa_N=11$ & 180& 445& 1.20& 1.58\\
\hline
\end{tabular} \end{center}
\end{table}

\begin{table}
\caption{Nucleon parameters in symmetric nuclear matter at the
phenomenological saturation value of nucleon density obtained by
employing of the nonrelativistic quark model. The values $(GA)$ in the
first and third lines correspond to the gas approximation. The values
in the second and fourth lines correspond to inclusion of the
three-body interactions, following Eq.\,(\ref{14n}). The results are
presented for two values of the parameter $\kappa_N$ --
Eq.\,(\ref{70h})}.
\begin{center}
\begin{tabular}{|c|c|c|c|c|} \hline
 &  $\Sigma_v$, MeV & $m^*$, MeV & $\lambda^2_m\rm\,,GeV^6$ &
$W^2_m\rm\,, GeV^2$ \\
\hline

$\rho=\rho_0$, $\kappa_N=8$ (GA)& 305& 685& 1.60&1.89\\

 $\rho=\rho_0$, $\kappa_N=8$ & 284 & 702& 1.84& 2.00\\
\hline

$\rho=\rho_0$, $\kappa_N=11$ (GA)& 291& 504& 1.11&1.59 \\

 $\rho=\rho_0$, $\kappa_N=11$ & 253& 530& 1.39& 1.71\\

\hline
\end{tabular} \end{center}
\end{table}

\clearpage

\newpage

\section*{Figure Captions}

FIG. 1. Contribution of the four-quark condensates (dark blob) to the
left-hand side of the sum rules caused by the valence quarks (dotted
lines). The helix line stands for the nucleon current. Solid lines
denote the quarks of the current.

FIG. 2. Contribution of the four-quark condensates (dark blob) to the
left-hand side of the sum rules caused by the sea quarks. Wavy line
stands for the pions. The other notations are the same as in Fig.\,1.
The $\pi QQ$ interaction vertices $1$ and $2$ may correspond to the
same or different nucleon quarks.

FIG. 3. Contribution of the four-quark condensates (dark blob) to the
left-hand side of the sum rules caused by the interference terms with
Fig.\,3a and Fig.\,3b illustrating the contact and vertex interference
correspondingly. The notations are the same as in Fig.\,2. The vertices 1, 2 and 3 may belong
to the same or
different nucleon quarks.

FIG. 4. Contribution of the four-quark condensates to the left-hand
side of the sum rules caused by the factorized terms. The small circles
denote the QCD vacuum. The dark blob stands for the two-quark
condensate.

FIG. 5. Density dependence of the nucleon vector self-energy $\Sigma_V$
and of the nucleon effective mass $m^*$ in the symmetric nuclear matter
for $\kappa_N=8$ corresponding to Eq.\,(79). The horizontal axis
corresponds to the density $\rho$ related to its saturation value
$\rho_0$. The dashed lines show the results in the gas approximation.
The solid lines correspond to inclusion of the most important 3-body
forces.

FIG. 6. Density dependence of the nucleon residue $\lambda_m^2$  and of
the effective threshold $W_m^2$ for $\kappa_N=8$ in the symmetric
nuclear matter corresponding to Eq.\,(79). The horizontal axis
corresponds to the density $\rho$ related to its saturation value
$\rho_0$. The other notations are the same as in Fig.\,5.

FIG. 7. Density dependence of the nucleon vector self-energy $\Sigma_V$
and of the nucleon effective mass $m^*$ in the symmetric nuclear matter
corresponding to Eq.\,(79) with inclusion of the 3-body forces. The
horizontal axis corresponds to the density $\rho$ related to its
saturation value $\rho_0$. The solid and dot--dashed lines show the
results for $\kappa_N=8$ and $\kappa_N=11$ correspondingly.

\newpage

\begin{figure}[ht]
\centerline{\epsfig{file=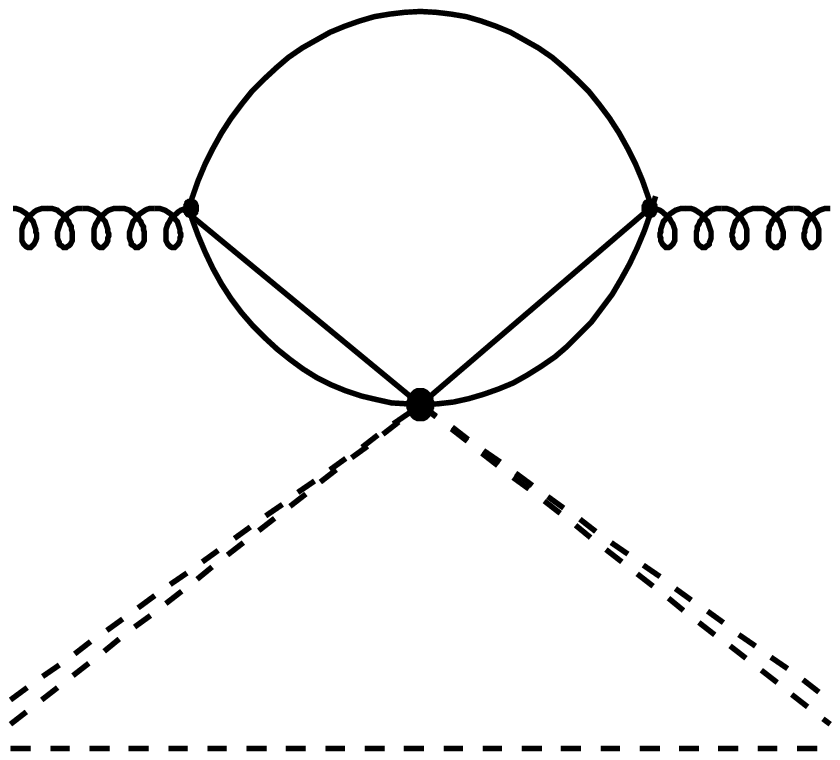,width=7.5cm}}
 \caption{}
 \end{figure}

\begin{figure}[ht]

\vspace{-0.5cm}

\centerline{\epsfig{file=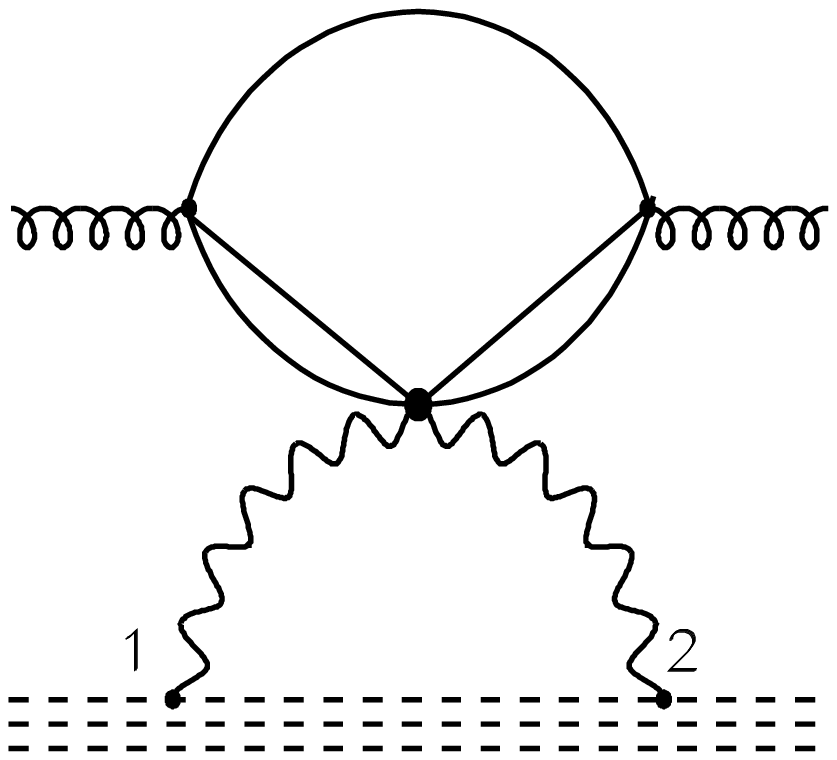,width=7.5cm}}
 \caption{}
 \end{figure}

\begin{figure}[ht]
\centerline{\epsfig{file=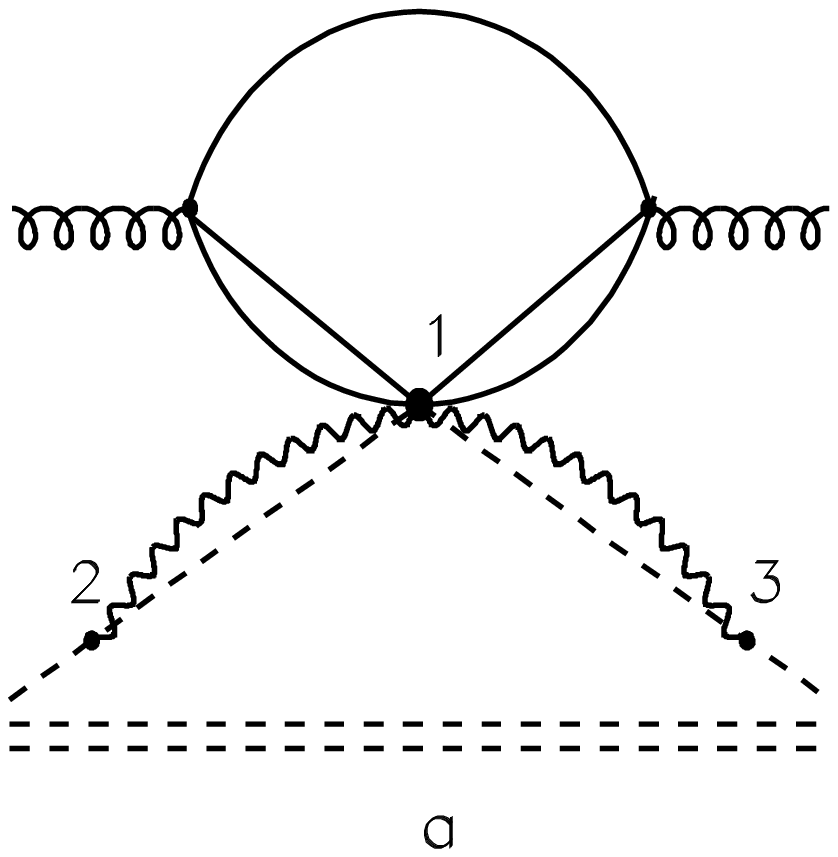,width=7.5cm}\epsfig{file=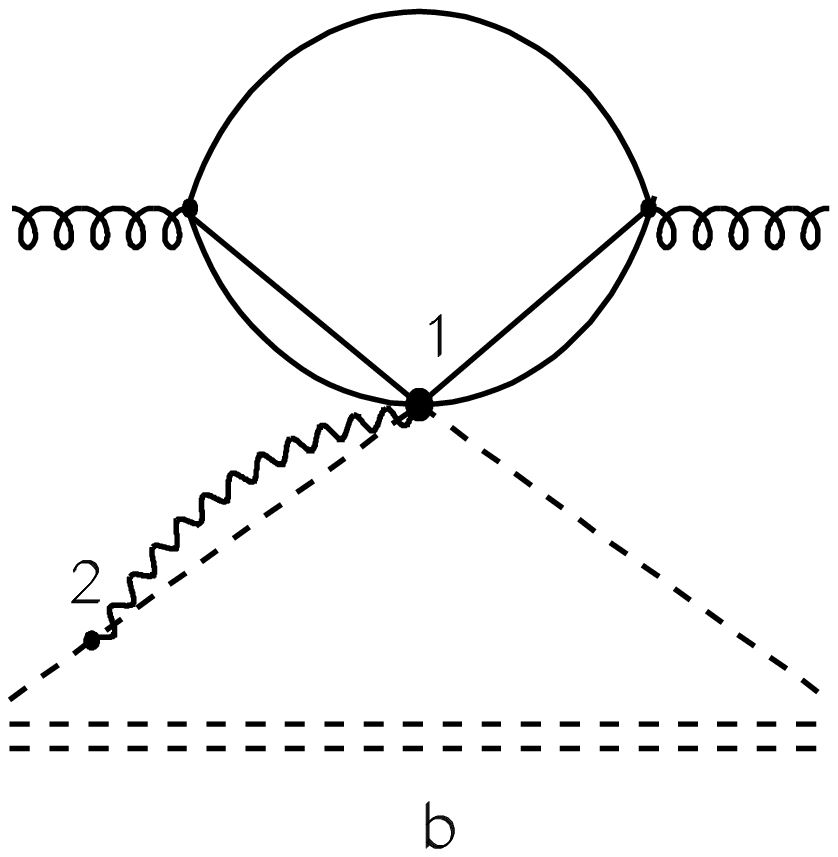,width=7.5cm}}
 \caption{}
 \end{figure}

\begin{figure}[ht]
\centerline{\epsfig{file=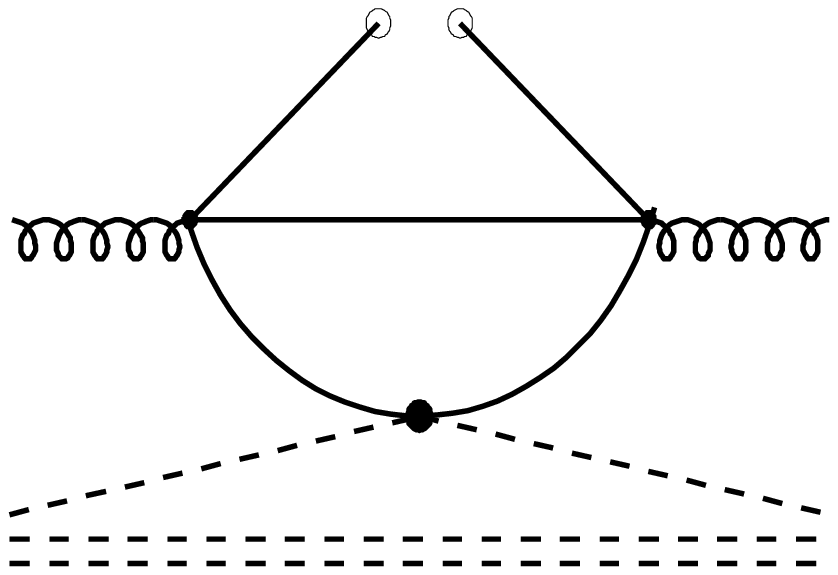,width=7.5cm}}
 \caption{}
 \end{figure}

\begin{figure}[ht]
\centerline{\epsfig{file=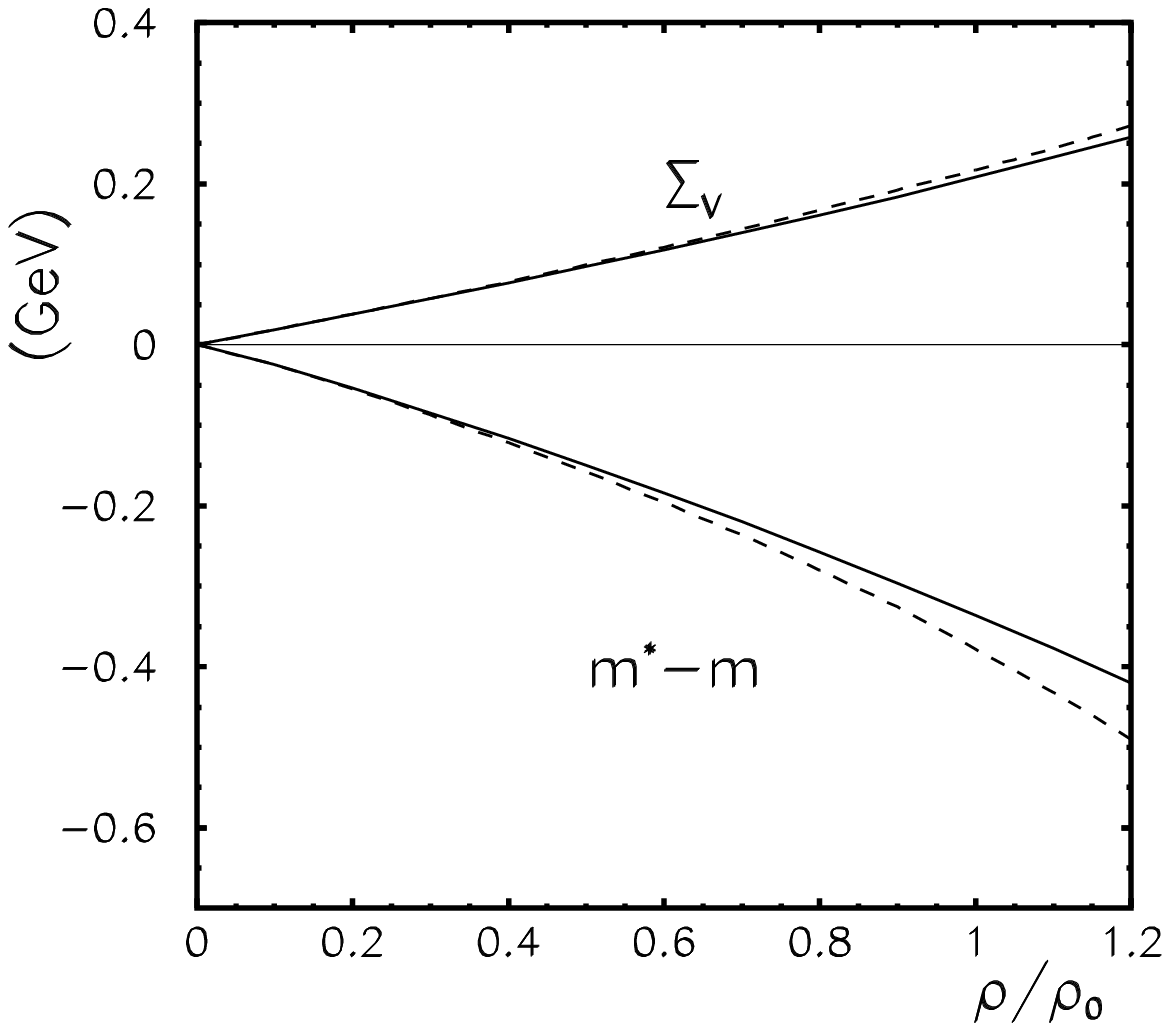,width=8.0cm}}
 \caption{}
 \end{figure}

\begin{figure}[ht]
\centerline{\epsfig{file=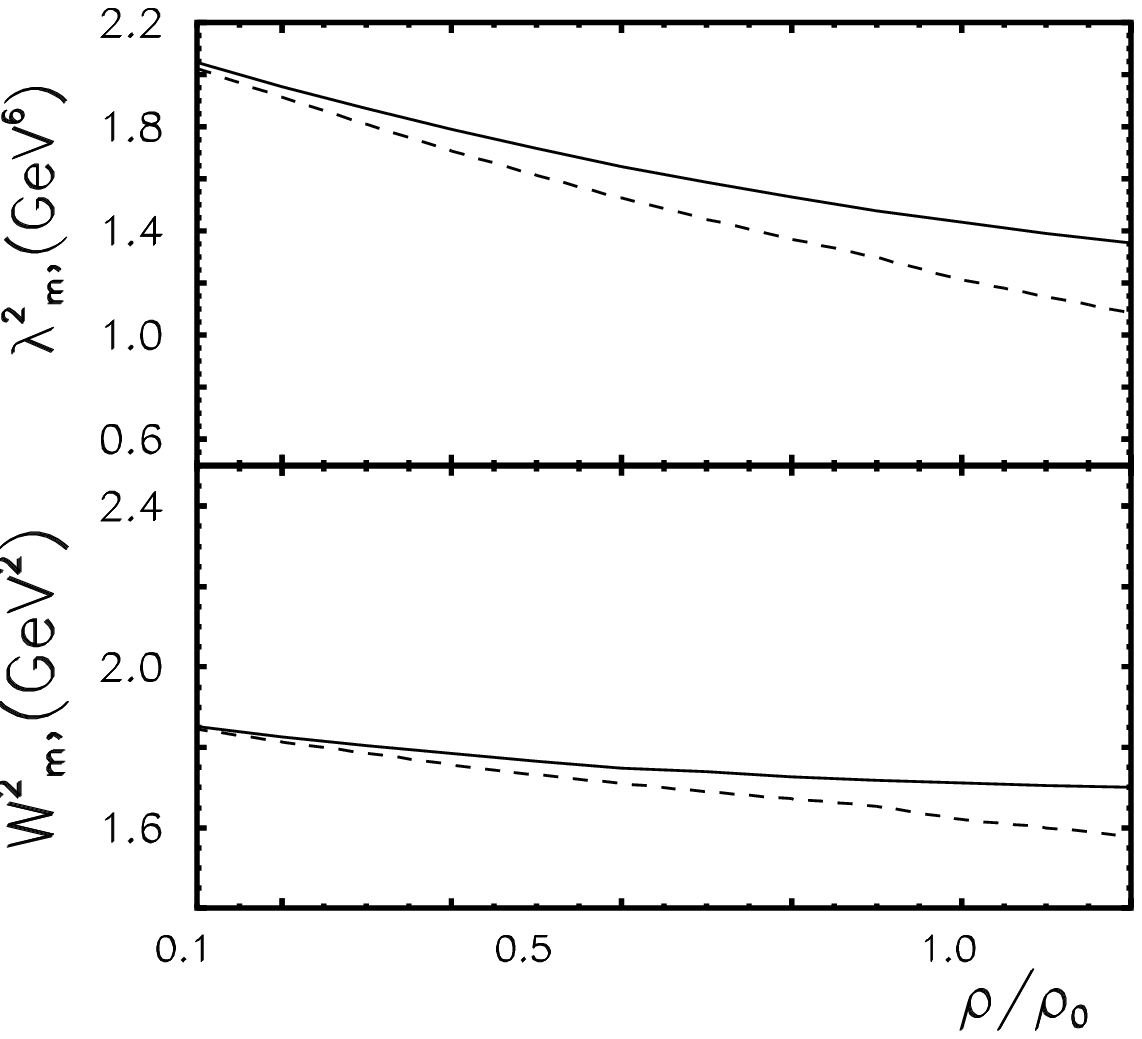,width=8.0cm}}
 \caption{}
 \end{figure}

\begin{figure}[ht]
\centerline{\epsfig{file=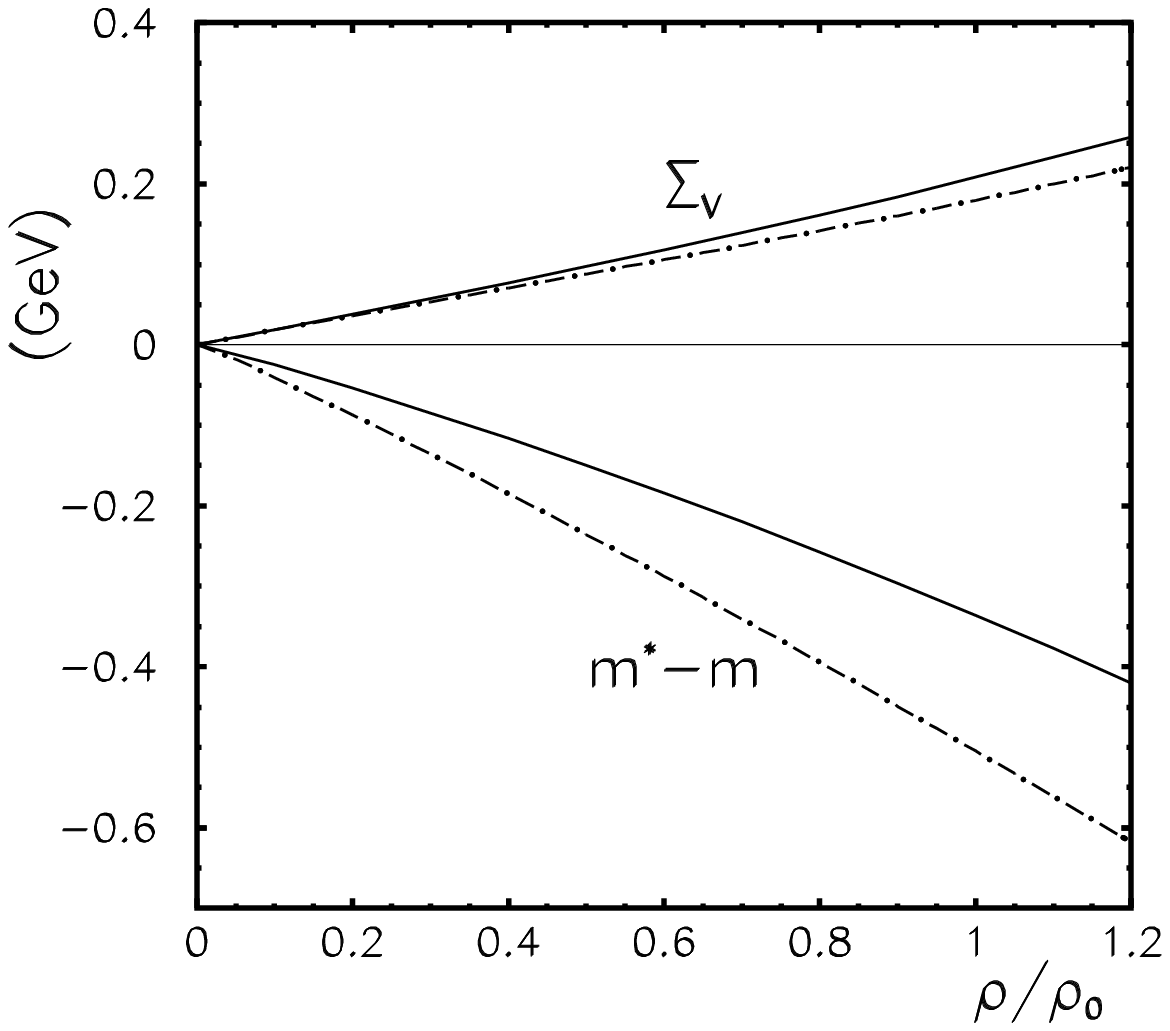,width=8.0cm}}
 \caption{}
 \end{figure}

\end{document}